\documentclass[unpublished]{JHEP3} 

\JHEPspecialurl{http://jhep.sissa.it/JOURNAL/JHEP3.tar.gz}

\usepackage{epsfig,multicol,bbm}

\newcommand\fverb{\setbox\fverbbox=\hbox\bgroup\verb}
\newcommand\fverbdo{\egroup\medskip\noindent%
                        \fbox{\unhbox\fverbbox}\ }
\newcommand\fverbit{\egroup\item[\fbox{\unhbox\fverbbox}]}
\newbox\fverbbox

\usepackage{graphicx}
\usepackage{amsmath}
\usepackage{amsfonts}
\usepackage{amssymb}
\def\effatlas{\eps_{ATLAS}}
\def\alam{A_{\lambda}}
\def\akap{A_{\kappa}}
\def\rts{\sqrt s}
\def\mupmum{\mu^+\mu^-}
\def\mmumu{M_{\mupmum}}
\def\upsi{\Upsilon_{1S}}
\def\mupsi{M_{\upsi}}

\def\upsiii{\Upsilon_{3S}}
\def\mupsiii{M_{\upsiii}}

\def\pbi{~\mbox{pb}^{-1}}

\def\cabb{C_{ab\anti b}}
\def\catt{C_{at\anti t}}
\def\camumu{C_{a\mu^-\mu^+}}
\def\catautau{C_{a\tau^-\tau^+}}

\def\cta{\cos\theta_A}
\def\ctamax{\cta^{\rm max}}
\def\sta{\sin\theta_A}

\def\ma{m_a}
\def\mh{m_h}

\def\hsm{h_{SM}}
\def\hi{h_1}

\def\ai{a_1}

\def\mhi{m_{h_1}}

\def\mai{m_{a_1}}

\def\mtau{m_\tau}

\def\beq{\begin{equation}}
\def\eeq{\end{equation}}
\def\bea{\begin{eqnarray}}
\def\eea{\end{eqnarray}}
\def\ie{{\it i.e.}}
\def\eg{{\it e.g.}}
\def\etc{{\it etc.}}
\def\lsim{\mathrel{\raise.3ex\hbox{$<$\kern-.75em\lower1ex\hbox{$\sim$}}}}
\def\gsim{\mathrel{\raise.3ex\hbox{$>$\kern-.75em\lower1ex\hbox{$\sim$}}}}
\def\ifmath#1{\relax\ifmmode #1\else $#1$\fi}

\def\fbi{~{\mbox{fb}^{-1}}}

\def\pb{~{\mbox{pb}}}
\def\br{BR}
\def\gev{~{\mbox{GeV}}}
\def\tev{~{\mbox{TeV}}}
\def\mev{~{\mbox{MeV}}}

\def\to{\rightarrow}

\def\calo{{\cal O}}

\def\dely{\Delta y}
\def\eps{\epsilon}

\def\anti{\overline}

    \def\fillboxx#1#2{\hbox to #1{\vbox to #2{\vfil}\hfil}   }

\def\tauptaum{\tau^+\tau^-}

\def\gev{~{\rm GeV}}
\def\gam{\gamma}
\def\tanb{\tan\beta}
\def\cotb{\cot\beta}

\def\anti{\overline}

\def\brupsimumu{\br(\upsi\to\mupmum)}
\def\epem{e^+e^-}

\newcommand{ \slashchar }[1]{\setbox0=\hbox{$#1$}   
   \dimen0=\wd0                                     
   \setbox1=\hbox{/} \dimen1=\wd1                   
   \ifdim\dimen0>\dimen1                            
      \rlap{\hbox to \dimen0{\hfil/\hfil}}          
      #1                                            
   \else                                            
      \rlap{\hbox to \dimen1{\hfil$#1$\hfil}}       
      /                                             
   \fi}     

\title{Direct production of a light CP-odd Higgs boson at the Tevatron
and LHC}

\author{Radovan Derm\'\i\v sek
   \\ Department of Physics, Indiana University, Bloomington, IN 47405, USA}

\author{John
  F. Gunion \\ Department of Physics, University of California, Davis,
  CA 95616, USA\\ and \\Theory Group, CERN, CH-1211, Geneva 23, Switzerland}

\abstract{We show that the existing CDF $L=630\pbi$ Tevatron data on
  $pp\to \mupmum X$ places substantial limits on a light CP-odd Higgs
  boson $a$ with $\ma<2m_B$ produced via $gg\to a$, even for
  $\ma>2\mtau$ for which $\br(a\to \mupmum)$ is relatively small.
  Extrapolation of this existing CDF analysis to $L=10\fbi$ suggests
  that Tevatron limits on the $ab\anti b$ coupling strength in the
  region $\ma>8\gev$ could be comparable to or better than limits from
  Upsilon decays in the $\ma<7\gev$ region. We also give rough
  estimates of future prospects at the LHC, demonstrating that early
  running will substantially improve limits on a light $a$ (or perhaps
  discover a signal). In particular, outside the Upsilon peak region,
  integrated luminosity of only $5\fbi-20 \fbi$ (depending on $\ma$ and
  $\rts$) could reveal a peak in $\mmumu$ and will certainly place
  important new limits on a light $a$.  The importance of such limits
  in the context of NMSSM Higgs discovery and $(g-2)_\mu$ are
  outlined.  } \received{} \accepted{}
\preprint{CERN-PH-TH/2009-196\\IUHET-536} \keywords{Higgs, Dimuon,
  Hadron Colliders}
\begin{document}

\section{Introduction}

Many motivations for the existence of a light CP-odd Higgs boson, $a$,
have emerged in a variety of contexts in recent years.  Of particular
interest is the $\ma<2m_B$ region, for which a light Higgs, $h$, with
SM-like $WW$, $ZZ$ and fermionic couplings can have mass below the
nominal LEP limit of $\mh>114\gev$ by virtue of $h\to aa\to 4\tau$
decays being
dominant~\cite{Dermisek:2005ar,Dermisek:2005gg,Dermisek:2006wr,Dermisek:2007yt}
(see also \cite{Chang:2005ht,Chang:2008cw}).  For $\mh\lsim 105\gev$,
the Higgs provides perfect agreement with the rather compelling
precision electroweak constraints, and for $\br(h\to aa)\gsim 0.75$
also provides an explanation for the $\sim 2.3\sigma$ excess observed
at LEP in $\epem \to Z b\anti b$ in the region $M_{b\anti b}\sim
100\gev$ if $\mh\sim 100\gev$.  This is sometimes referred to as the
``ideal'' Higgs scenario.  More generally, superstring modeling
suggests the possibility of many light $a$'s, at least some of which
couple to $\mupmum$, $\tauptaum$ and $b\anti b$.  Further, it is not
excluded that a light $a$ with $\ma>8\gev$ and enhanced $ab\anti b$
coupling could be responsible for the deviation of the measured muon
anomalous magnetic moment $a_\mu$ from the SM
prediction~\cite{Gunion:2008dg}. Below, we will show that a light $a$
with the required $ab\anti b$ and $a\mupmum$ couplings would have been
seen in existing Tevatron data for the $\mupmum$ final state at low
$\mmumu$. More generally, current muon pair Tevatron data places
significant limits on a light $a$. These will be further strengthened
with increased Tevatron integrated luminosity and by $\mupmum$ data
obtained at the LHC.

The possibilities for discovery of an $a$ and limits on the $a$ are
phrased in terms of the $a\mu^-\mu^+$, $a\tau^-\tau^+$, $ab\anti b$
and $at\anti t$ couplings defined via
\beq
{\cal L}_{af\anti f}\equiv i C_{af\anti f}{ig_2m_f\over2m_W}\anti f \gamma_5 f
a\,.
\label{cabbdef}
\eeq 
In this paper, we assume a Higgs model in which
$\camumu=\catautau=\cabb$, as typified by a two-Higgs-doublet model
(2HDM) of either type-I or type-II, or more generally if the lepton
and down-type quark masses are generated by the same combination of
Higgs fields. However, one should keep in mind that there are models
in which $r=(\camumu=\catautau) /\cabb\gg 1$ --- such models include
those in which the muon and tau masses are generated by different
Higgs fields than the $b$ mass.  In a 2HDM of type-II and in the MSSM,
$\camumu=\catautau=\cabb=\tanb$ (where $\tanb=h_u/h_d$ is the ratio of
the vacuum expectation values for the doublets giving mass to up-type
quarks vs.  down-type quarks) and $\catt=\cotb$.  These results are
modified in the NMSSM (see, \eg\ \cite{Ellis:1988er} and \cite{hhg}).
\footnote{A convenient program for exploring the NMSSM Higgs sector is
  NMHDECAY~\cite{Ellwanger:2004xm,Ellwanger:2005dv}.} In the NMSSM,
both $\catt$ and $\cabb=\camumu=\catautau$ are multiplied by a factor
$\cta$, where $\cta$ is defined by
\beq
a=\cta a_{MSSM}+\sta a_S,
\eeq
where $a$ is the lightest of the 2 CP-odd scalars in the model
(sometimes labeled as $\ai$).  Above, $a_{MSSM}$ is the CP-odd
(doublet) scalar in the MSSM sector of the NMSSM and $a_S$ is the
additional CP-odd singlet scalar of the NMSSM.  In terms of $\cta$,
$\camumu=\catautau=\cabb=\cta\tanb$ and $\catt=\cta\cotb$.  Quite
small values of $\cta$ are natural when $\ma$ is small as a result of
being close to the $U(1)_R$ limit of the model.  In the most general
Higgs model, $\camumu$, $\catautau$, $\cabb$ and $\catt$ will be more
complicated functions of the vevs of the Higgs fields and the
structure of the Yukawa couplings. In this paper, we assume
$\camumu=\catautau=\cabb$ and $\cabb/\catt=\tan^2\beta$. 

One should keep in mind, however, the fact that the above are
tree-level couplings and that the $b\anti b a_{MSSM}$ coupling is
especially sensitive to radiative corrections from SUSY particle loops
that can be large when $\tanb$ is
large~\cite{Hall:1993gn,Carena:1994bv,Pierce:1996zz}. These are
typically characterized by the quantity $\Delta_b$ which is crudely of
order ${\mu\tanb\over 16\pi^2M_{SUSY}}$. The correction to the coupling
then takes the form of $1/(1+\Delta_b)$.  Since $\mu$ can have either
sign, $\cabb$ can be either enhanced or suppressed relative to
equality with $\catautau $ (the corrections to which are much smaller)
and $\camumu$ (the corrections to which are negligible).

In the past, probes of a light $a$ have mainly relied on production of
a primary particle (\eg\ an Upsilon) which then decays to a lighter
$a$ with the emission of a known SM particle (\eg\ a photon). Such
probes are strictly limited to a maximum accessible $\ma$ by simple
kinematics. The only exceptions to this statement have been probes
based on $\epem\to b\anti b a$ followed by $a\to\tauptaum$ or $a\to
b\anti b$, with LEP providing the strongest (but still rather weak)
limits on the $a$ based on this type of radiative production process.

In contrast, hadron colliders potentially have a large reach in $\ma$
as a result of the fact that an $a$ can be produced via $gg\to a$. The
$gga$ coupling derives from quark triangle loops. This process, plus
higher order corrections thereto, leads to a large cross section for
the $a$ due to the large $gg$ ``luminosity'' at small gluon momentum
fractions, provided the $aq\anti q$ coupling deriving from the doublet
component of the $a$ is significant.  This large cross
section will typically lead to a significant number of $gg\to a \to
\mupmum$ events even though $\br(a\to\mupmum)$ is not very large (and
in fact is quite small for $\ma>2\mtau$).  Further, since the $a$ is a
very narrow resonance, all $a$ events will typically fall into a
single bin of size given by the $\mmumu$ mass resolution of the
experiment, typically below $100\mev$. This implies very controllable
background levels, mostly deriving from heavy flavor production (\eg\
$b\anti b$ and $c\anti c$), once isolation and promptness cuts have
been imposed on the muons.

The Higgs doublet component of the $a$ can be suppressed when the $a$
mixes with one or more SM singlet CP-odd field, \eg\ the $a_S$ of the
NMSSM as made explicit above. However, what is important for limits is
$\cabb$.  In the NMSSM context, the scenarios allowing a light scalar
Higgs to escape LEP limits by virtue of $h\to aa$ decays with
$\ma<2m_B$ are such that $\cta$ is only small when $\tanb$ is large;
for example, preferred scenarios for $\tanb=10$ are such that
$\cta\sim 0.1$, implying $\cabb\sim 1$.  As a result, the Tevatron and
LHC provide very significant probes of such a light $a$ despite its
being only 1\% doublet (at the probability level). In addition, there
are models beyond the MSSM~\cite{Dermisek:2008id,Dermisek:2008sd},
including scenarios within the NMSSM~\cite{Dermisek:2008uu}, in which
the doublet component of the $a$ can be quite substantial. (For a
review, see also \cite{Dermisek:2009si}.) In such models there are
typically several Higgs scalars (including charged Higgs bosons) with
masses near $100\gev$ that have escaped discovery because of decays
involving the light $a$. This type of scenario requires that $\tanb$
be small ($1\lsim \tanb\lsim 3$). When the doublet component of the
$a$ is substantial $\cabb=\cta\tanb$ will have magnitude $\geq 1$ and
hadron colliders will almost certainly discover or exclude the $a$.
In the case of the NMSSM, there is a portion of the preferred
parameter region for low $\tanb$ in which precisely this kind of
scenario arises. However, in the NMSSM at low $\tanb$ there is a
second part of the preferred parameter region in which there are many
light Higgs bosons but the $a$ is mainly singlet.  In this latter
case, $|\cabb|$ will be relatively small and direct searches for the
light $a$ will be more difficult than in NMSSM models with larger $\tanb$.

The organization of the paper is as follows. In
Sec.~\ref{currentlimits}, we review some basic facts about a light $a$
and limits on $\cabb=\camumu=\catautau$ coming from
non-hadron-collider data.  In Sec.~\ref{tevatron}, we discuss the
additional limits that can be placed on the couplings of the $a$
implied by existing Tevatron analyses and data and extrapolate these
existing results to $L=10\fbi$ data sets.  In Sec.~\ref{lhc}, we
analyze prospects for discovering, or at least further improving
limits on the couplings of, a light $a$ using early LHC data.
Sec.~\ref{conclusions} summarizes our conclusions and provides a few
additional comments.

\section{Phenomenology and limits for a light CP-odd $a$} 
\label{currentlimits}

\begin{figure}[b!]
\begin{center}
\includegraphics[width=0.65\textwidth,angle=90]{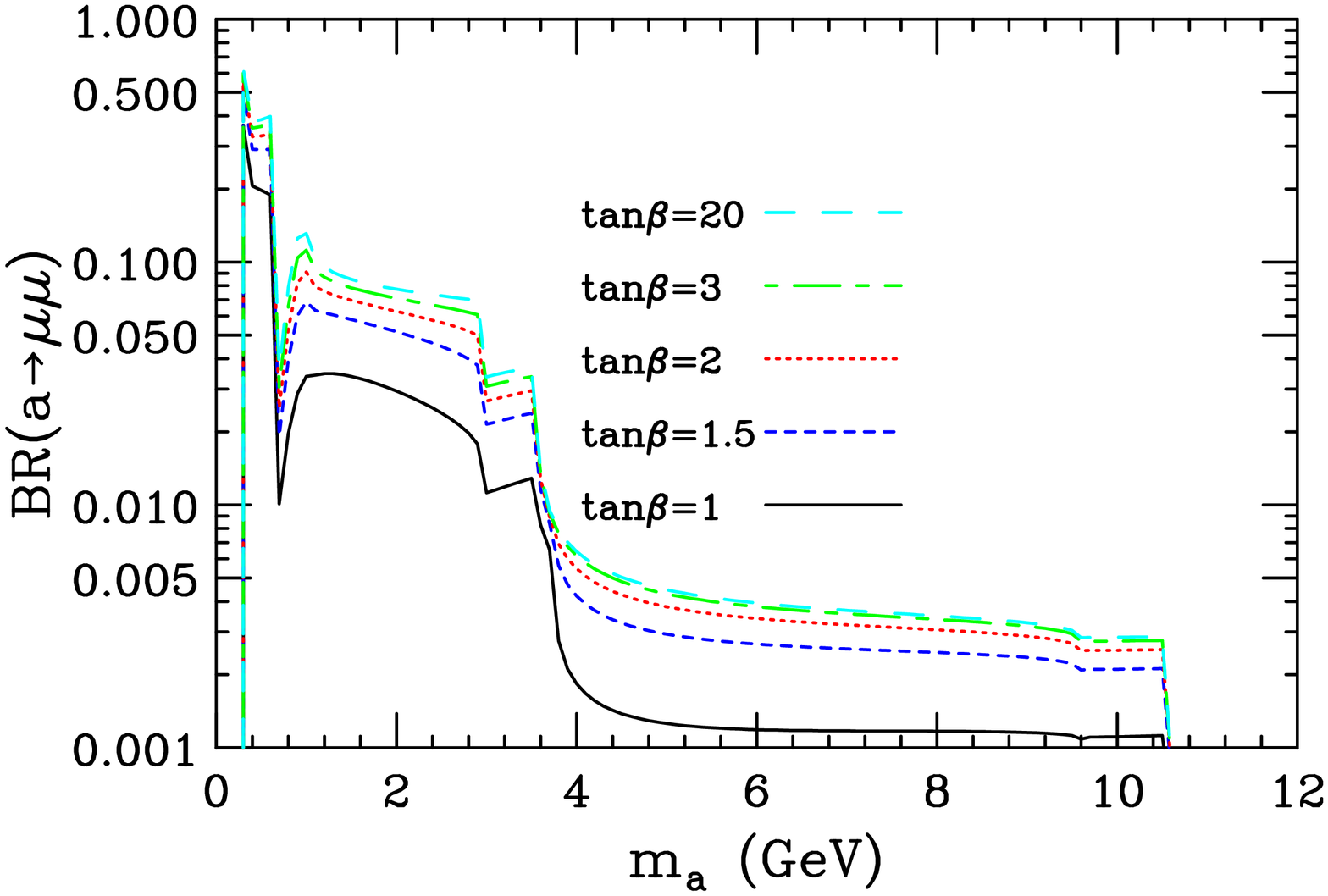}
\end{center}
\caption{$\br(a\to \mupmum)$ is plotted as a function of $\ma$ for a
  variety of $\tanb$ values. $\br(a\to\mupmum)$ is independent of
  $\cta$ at tree-level.
}
\label{bramumu}
\end{figure}
One key ingredient in understanding current limits and future
prospects is the branching ratio for $a\to \mupmum$ decays. This
branching ratio (which is independent of $\cta$ at tree-level due to
the absence of tree-level $a\to VV$ couplings and similar) is plotted
in Fig.~\ref{bramumu}. Note that $\br(a\to\mupmum)$ changes very
little with increasing $\tanb$ at any given $\ma$ once $\tanb\gsim 2$.

\begin{figure}[h!]
\begin{center}
\includegraphics[width=0.65\textwidth,angle=90]{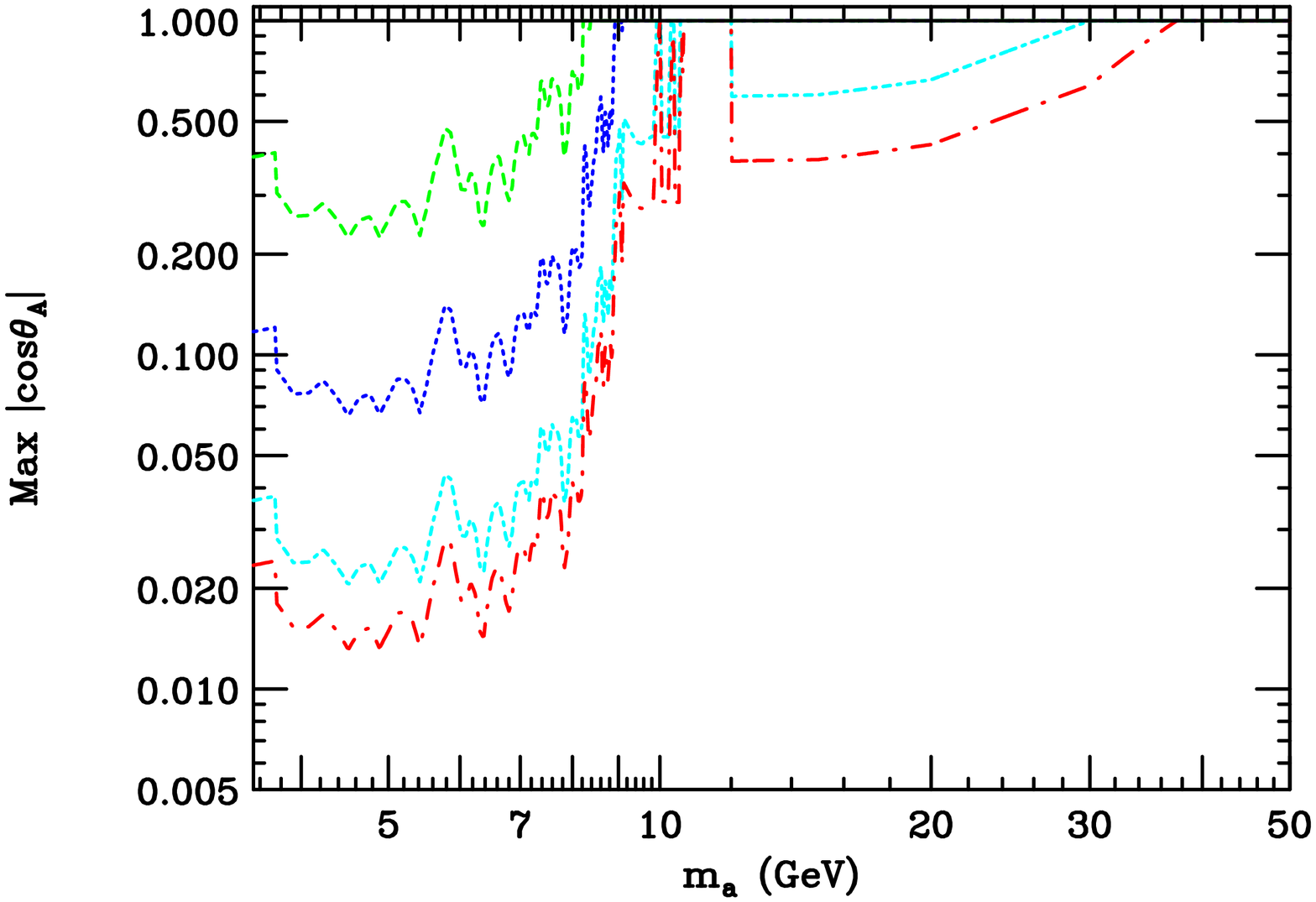}
\end{center}
\caption{We plot results from \cite{Gunion:2008dg} for $\ctamax$ in
  the NMSSM (where $\cabb=\cta\tanb$) as a function of $\ma$ for
  $\ma>2\mtau$.  The different curves correspond to $\tanb=1$ (upper
  curve), $3$, $10$, $32$ and $50$ (lowest curve). The region between
  $\sim 2\mtau$ and $\sim 8\gev$ is strongly constrained by CLEO III
  data \cite{:2008hs} on $\upsi\to \gam \tauptaum$ decays.  The
  plotted limits do not include the BaBar $\upsiii\to \gam \tauptaum$
  and $\upsiii\to \gam\mupmum$ limits that became available after the
  analysis of \cite{Gunion:2008dg}.}
\label{ctamaxvsma}
\end{figure}

Limits on $|\cabb|=|\cta|\tanb$ were analyzed in \cite{Gunion:2008dg}
(see also \cite{Domingo:2008rr}), based on data available at the
time. The analysis of \cite{Gunion:2008dg} employed limits from
$\Upsilon\to \gam a$ decays, the importance of which was emphasized in
\cite{Dermisek:2006py} (especially within the NMSSM context), as well
as from $\epem\to b\anti b a$ production at LEP. The analysis of
\cite{Gunion:2008dg} was done prior to the very recently released
BaBar $\Upsilon(nS)\to \gam a$ results
\cite{Aubert:2009ck,Aubert:2009cp}.  Without including the $\upsiii$
BaBar data, limits in the $8\gev<\ma<2m_B$ range (especially,
$\mupsi<\ma<2m_B$) are quite weak and suffer from uncertainty
regarding $\eta_b-a$ mixing. An update employing the $\upsiii$ data
will be performed in a separate paper.  In the present paper, the limits
implicit in Tevatron data are compared to the limits obtained in
\cite{Gunion:2008dg}. We will also briefly summarize how this
comparison will change after inclusion of the $\upsiii$ BaBar results.

Focusing on the NMSSM, we note that it is always possible to
choose $\cta$ so that the limits on $\cabb$ as a function of $\tanb$
are satisfied. The maximum allowed value of $|\cta|$, $\ctamax$, as a
function of $\ma$ for various $\tanb$ values as obtained in
\cite{Gunion:2008dg} is plotted in Fig.~\ref{ctamaxvsma}. Constraints
are strongest for $\ma\lsim 7\gev$ for which Upsilon limits are
strong, and deteriorate rapidly above that.

Turning to the 2HDM(II), where $\cabb=\tanb$, we note that any point
for which $\ctamax$ is smaller than $1$ corresponds to an $\ma$ and
$\tanb$ choice that is not consistent with the experimental limits.
Disallowed regions emerge for $\ma\lsim 10\gev$ at higher $\tanb$.  A
disallowed region also arises over a limited $\ma$ range starting from
$\ma>12\gev$ when $\tanb\gsim 18$, the larger the value of $\tanb$ the
larger the interval.  For example, for $\tanb=50$ the 2HDM(II) is not
consistent for $\ma<10\gev$ nor for $12\lsim \ma\lsim 37\gev$. In
contrast, for $\tanb=10$ the 2HDM(II) model is only inconsistent for
$\ma\lsim 9\gev$.

Before proceeding, we note that constraints from precision electroweak
data are easily satisfied for a light $a$ in both the 2HDM(II) and
NMSSM cases (see~\cite{Gunion:2008dg} for more discussion).  We also
wish to make note of the regions of interest for obtaining a new
physics contribution, $\Delta a_\mu$, of order $\Delta a_\mu\sim
27.5\times 10^{-10}$ (the current discrepancy between observation and
the SM prediction). These can be roughly described as follows.  In the
2HDM(II) context, such $\delta a_\mu$ requires a rather precisely
fixed value of $\tanb\sim 30-32$ and $\ma\sim 9.9-12\gev$.  In the
NMSSM context, the strong constraints from Upsilon physics imply that
significant contributions to $a_\mu$ are not possible until $\ma$
exceeds roughly $9.2\gev$. The maximal $\delta a_\mu$ can exceed
$\Delta a_\mu\equiv 27.5\times 10^{-10}$ for $9.9\gev\lsim \ma\lsim
12\gev$ if $\tanb\geq 32$, with an almost precise match to this value
for $\tanb=32$.  For $\tanb=50$, one can match $\Delta a_\mu$ by using
a value of $\cta$ below $\ctamax$. (The fact that matching is possible
for $9.9\gev\lsim \ma\lsim 2m_B$ is particularly interesting in the
context of the ideal Higgs scenario.) Further, the maximal $\delta
a_\mu$ is in the $7-20\times 10^{-10}$ range for $12\gev < \ma\lsim
48\gev$ for $\tanb=32$ and for $12\gev < \ma\lsim 70\gev$ for
$\tanb=50$.

At this point, it is worth discussing in more depth the ``ideal''
$\mh\sim 100\gev$, $\ma\lsim 2m_B$, $\br(h\to aa)>0.75$ Higgs scenario
as discussed
in~\cite{Dermisek:2005ar,Dermisek:2005gg,Dermisek:2006wr,Dermisek:2007yt}.
These references examined the degree to which obtaining the observed
value of $m_Z$ requires very precisely tuned values of the GUT scale
parameters of the MSSM and NMSSM.  One finds that in any
supersymmetric model this finetuning is always minimized for GUT scale
parameters that yield a SM-like $h$ with $\mh\leq 100-105\gev$, something
that is only consistent with LEP data if the $h$ has unexpected decays
that reduce the $h\to b\anti b$ branching ratio while not contributing
to $h\to b\anti b b\anti b$ (also strongly constrained by LEP data).
A Higgs sector with a light $a$ for which $\br(h\to aa)>0.75$ and with
$\ma$ small enough that $a$ decays to $B\anti B$ final states are
disallowed (i.e. $\ma < 10.56\gev$) provides a very natural
possibility for allowing minimal finetuning. The NMSSM provides one
possible example. As a useful benchmark, in the context of the NMSSM
the $\tanb=10$ scenarios that yield the required $\ma<2m_B$ and
$\br(h\to aa)>0.75$ are ones with $0.35\gsim |\cabb|$ ($|\cta|\gsim
0.035$).  The lower limit arises from the fact that $\br(h\to aa)$
falls below the $0.75$ level needed for the ideal Higgs scenario if
$|\cta|$ is too small. From Fig.~\ref{ctamaxvsma} we see that such
$|\cta|$ values are not yet excluded for any $\ma>2\mtau$.  This range
becomes more restricted if, in addition, one requires small finetuning
of the $\alam$ and $\akap$ soft-SUSY-breaking NMSSM parameters that
determine the properties of the $a$ --- such finetuning is
characterized by a parameter we call $G$, defined in
\cite{Dermisek:2006wr}.  At $\tanb=10$, $0.6\lsim |\cabb|\lsim 1.2$
($0.06\lsim |\cta|\lsim 0.12$) is required if $G<20$ is imposed as
well as requiring $\ma<2m_B$ and $\br(h\to aa)>0.75$.  For $\tanb\lsim
2$, the means for escaping the LEP constraints on the light scalar
Higgses are a bit more complex since two $\sim 100\gev$ Higgses
can share the $ZZ$-Higgs couplings squared, but there is always a
lower limit on $|\cta|$ for which such escape is possible. In
Table~\ref{ctatable}, we tabulate more precisely the values of $\cta$
for various $\tanb$ values that: a) have $\ma<2m_B$ and large enough
$\br(h\to aa)$ to escape LEP limits on the $h$, with no constraint on
$G$; and b) have small $\akap,\alam$ finetuning measure $G<20$ as well
as $\ma<2m_B$ and large enough $\br(h\to aa)$.
\begin{table}
  \caption{Values of $\cta$ required for $\ma<2m_B$ and sufficiently
    large $\br(h\to
    aa)$ to escape LEP limits on the $Zb\anti b$ final state. Results
    both without and with $G<20$ required are presented for a selection of
    $\tanb$ values.\label{ctatable}}
\smallskip
\begin{center}
\begin{tabular}{|c|c|c|}
\hline
$\tanb$ & $\cta$ ranges & $\cta$ ranges, $G<20$ required\cr
\hline
1.7 & $<-0.3$ or $>0.1$ & $[-0.6,-0.5]$ or $\sim +0.1$ \cr
2 & $<-0.3$ or $>0.1$ & $[-0.7,-0.5]$ \cr
3 & $<-0.06$ & $[-0.35,-0.08]$ \cr
10 & $<-0.06$ or $>0.035$ & $[-0.12,-0.08]$ or $[0.06,0.08]$ \cr
50 & $<-0.04$ or $>0.04$ & $[-0.06,-0.04]$ or $\sim +0.04$ \cr
\hline
\end{tabular}
\end{center}
\end{table}
 
We can summarize the implications of this table as follows.  First,
comparing to the existing limits on $|\cta|$ as plotted in
Fig.~\ref{ctamaxvsma}, we see that only ideal Higgs scenarios (\ie\
ones with $\mh<105\gev$ and $\br(h\to aa)$ large enough to escape LEP
limits) with $\tanb>30$ and $\ma\lsim 8\gev$ are excluded. Ideal Higgs
scenarios with $\tanb<10$ are fairly far from being excluded. If we
wish to eliminate ideal Higgs scenarios then: for $1.7\lsim\tanb\lsim
2$,\footnote{Scenarios with $\tanb$ much below $1.7$ are problematical
  since it is difficult to retain perturbativity for Yukawa couplings
  all the way up to the unification scale.} we must exclude
$|\cabb|\gsim 0.17$; for $\tanb=3$, we must exclude $|\cabb|\gsim
0.18$; for $\tanb=10$, we must exclude $|\cabb|\gsim 0.35$ and for
$\tanb=50$, we must exclude $|\cabb|\gsim 2$. If we only wish to
exclude such scenarios that also have $G<20$, then the required
$|\cabb|$ levels for $\tanb=1.7,2,3,10,50$ are $0.17, 1, 0.24, 0.6,2$,
respectively.  As we shall see, completely probing even the latter
levels for all $\ma<2m_B$  will be challenging,
but hadron colliders may ultimately play a leading role.  Indeed,
those scenarios with $G<20$ typically have $\ma$ values above
$7.5\gev$ and most often above $\mupsiii$. Of course, the many scenarios
with larger $|\cta|$ than the values listed above will be
correspondingly easier to exclude or verify.

Finally, we comment on the implications of the recent 
ALEPH results \cite{cranmer} which place a limit on
\beq
\xi^2\equiv {\sigma(\epem \to Z \hi)\over \sigma(\epem\to Z
  \hsm)}\br(\hi\to \ai\ai)[\br(\ai\to \tauptaum)]^2\,
\eeq
as a function of $\mhi$ and $\mai$.\footnote{In this arXiv version of
  this paper, we have access to final results and so we have updated
  this paragraph relative to the published version of this paper.}
(Above, we use the more precise notation $\hi$ and $\ai$ for the
lightest CP-even and CP-odd Higgs bosons of the NMSSM.)  
According to the ALEPH analysis, to
have $\mhi\lsim 100\gev$, $\xi^2_1\lsim 0.52$ ($0.42$) is required if
$\mai\sim 10\gev$ ($4\gev$).  These limits rise rapidly with
increasing $\mhi$ --- for $\mhi=105\gev$ (the rough upper limit on
$\mhi$ such that electroweak finetuning remains quite small and
precision electroweak constraints are fully satisfied) the ALEPH
analysis requires $\xi^2\lsim 0.85$ ($\lsim 0.7$) at $\mai\sim
10\gev$ ($4\gev$).  These limits are such that the easily viable NMSSM
scenarios are
ones: i) with $\mai$ below but fairly close to $2m_B$,
which is, in any case, strongly preferred by minimizing the
light-$\ai$ finetuning measure $G$; and/or ii) with $\tanb$ relatively
small ($\lsim 2$).
These are also the scenarios for which Upsilon constraints are either
weak or absent. Details will be provided
in a forthcoming paper~\cite{upsilonupdate}. Here, we present a simple
summary. In particular, we note the following: a)~all
$\tanb\leq 2$ cases provide $\mhi\leq 100\gev$ scenarios that escape the
ALEPH limits; b)~there are a few $G<20$, $\tanb=3$ scenarios with
$\mhi$ as large as
$98\gev$ and $99\gev$ and with $\xi^2$  essentially equal to
the ALEPH limits of $\xi^2\leq 0.42$ and $\xi^2\leq 0.45$ applicable at
these respective $\mhi$ values; c)~$\tanb=10$ ideal scenarios easily
allow for $\mhi\sim 100-105\gev$ (because the tree-level Higgs mass is
larger at $\tanb=10 $ than at $\tanb=3$) and at $\mai\lsim 2m_B$ many
$\mhi\gsim 100\gev$ points have
$\xi^2<0.5$ in the fixed-$\mu$ scan and a few of the full-scan points
have $\xi^2<0.6$ for $\mhi\sim 105\gev$, both of which are below the
$\mai=10\gev$ ALEPH upper limits on $\xi^2$ of 0.52 at $\mhi\sim 100\gev$ and
$0.85$ at $\mhi=105\gev$; d)~at
$\tanb=50$ there are some $G<20$ points with $\mhi\sim 100\gev$ and
$\mai\lsim 2m_B$ having $\xi^2$ below the $0.52$ ALEPH limit. 
Finally, we note that for the entire range of Higgs masses
studied the ALEPH limits were actually $\sim 2\sigma$ stronger than
expected. Thus, it is not completely unreasonable to consider the
possibility that the weaker expected limits should be employed.  These
weaker limits for example allow $\xi^2$ as large as $0.52$ at $\mhi\sim
95\gev$ and $0.9$ for $\mhi\sim 100\gev$. These weaker limits allow
ample room for the majority of the $\mai\lsim 2m_B$ ideal Higgs scenarios.

\section{The role of the Tevatron}
\label{tevatron}

Potentially, the Tevatron can probe precisely the $\ma$ range close to
and above the $\Upsilon(nS)$ masses which cannot be probed in
$\Upsilon(nS)$ decays.  Some relevant analyzes have been performed
looking for a very narrow resonance, denoted $\eps$, that is produced
in the same way as the $\upsi$.  The published/preprinted results are
those of \cite{Apollinari:2005fy} and \cite{Aaltonen:2009rq} from the
CDF experiment.  The latter results employ data corresponding to
$L=630\pbi$ and exclude the potential $\eps$ peak at $\mmumu\sim
7.2\gev$ present in the $L=110\pbi$ data of the first paper.  The
analysis was only performed for the region from $6.3\gev\leq \mmumu
\leq 9\gev$. The reason for not performing the analysis at lower
$\mmumu$ is that the acceptance of the $\mupmum$ pair relative to that
of the $\upsi$ (used as a normalizing cross section) would be highly
mass dependent.  This is due to the fact that CDF is only able to see
$\mupmum$ pairs with $p_T>5\gev$ and for $\mmumu<6.3\gev$ the fraction
of pairs that fail this cut becomes highly mass dependent.  No reason
for not analyzing the region of $\mmumu>9\gev$ is given, although it
is in this region that the $\Upsilon(1S,2S,3S)$ peaks are present.

Our goal here is to use the above $\eps$ analyses to place limits on a
CP-odd $a$. This is possible under certain assumptions detailed below.
In particular, we will place limits on $\cabb$.

\begin{figure}
\begin{center}
\includegraphics[width=0.65\textwidth,angle=90]{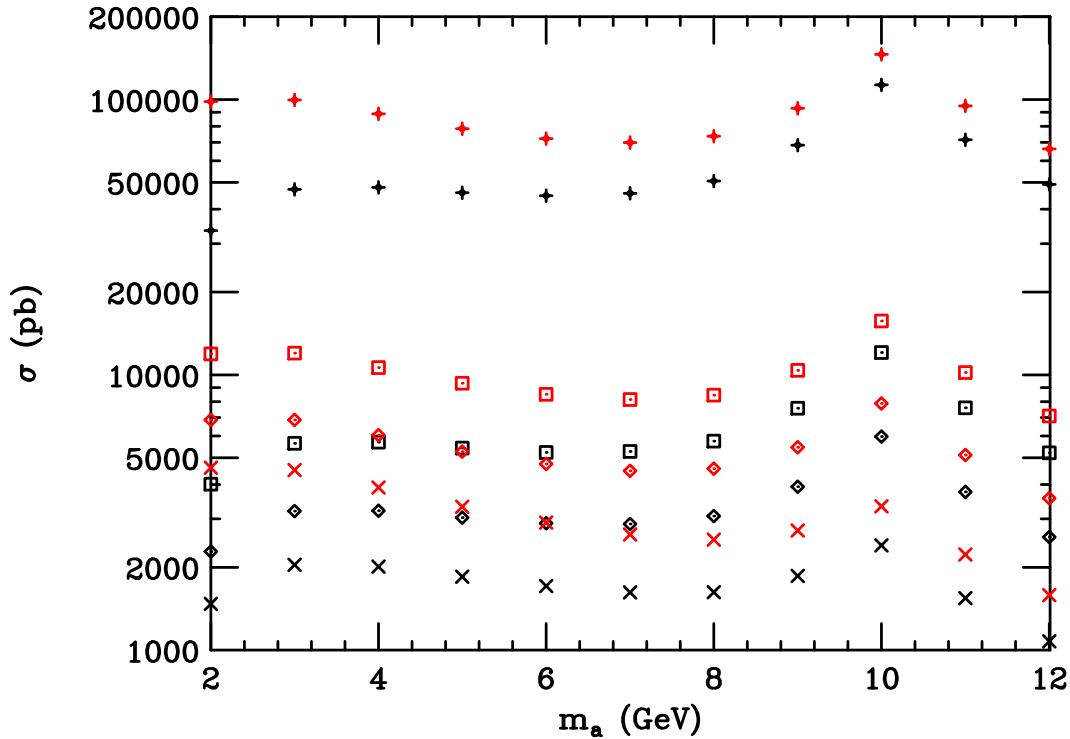}
\end{center}
\caption{The total cross section for $a$ production at the Tevatron is plotted vs
  $\ma$ for $\tanb=1,2,3,10$ (lowest to highest point sets).  For each
  $\ma$ and $\tanb$ value, the lower (higher) point is the cross
  section without (with) resolvable parton final state contributions.
}
\label{totsigs}
\end{figure}

The dominant production mechanism for a light $a$ at a hadron collider
is different than that for the $\eps$ (assumed to be the same or very
similar in kinematic shape \etc\ for the $\upsi$).  The production
mechanisms for the $\upsi$ remain uncertain. It has recently been
claimed \cite{Artoisenet:2008fc} that an NNLO version of the leading
order (LO) calculation can reproduce the Tevatron results for direct
production of the $\upsi$ at larger $p_T$ (roughly $p_T>5\gev$).  The
diagrams employed begin with the LO $\calo(\alpha_S^3)$ process $gg\to
b\anti b g$ with the $b\anti b$ pair turning into the $\upsi$ with
probability determined by $|R_{\upsi}(0)|^2$, leading to an $\upsi+g$
final state.  At NLO, the $\alpha_S^4$ diagrams include virtual
correction diagrams that also lead to the $\upsi+g$ final state and
several diagrams containing an extra quark or gluon in the final state
($\upsi+2g$ and $\upsi+b\anti b$). Several $\calo(\alpha_S^5)$
diagrams leading to $\upsi+3j$ final states (especially $\upsi+3g$)
are argued to be of importance at larger $p_T$ and are also included.
Resummation \cite{Berger:2004cc} is necessary to get the low $p_T$
portion of the cross section.  From CDF and D0 data, the direct
$\upsi$ production cross section is measured to be about 50\% of the
total. Indirect contributions coming from, for example, $gg\to \chi_b$
followed by $\chi_b\to \upsi \gamma$ make up the remaining 50\% of the
total $\upsi$ production rate.  In contrast, $a$, being a spin-0
resonance, will be dominantly produced via $gg\to a$ through the
quark-loop induced $gga$ coupling.  In addition, there are large QCD
corrections to the one-loop-induced cross section.  These are of two
basic types: a) virtual corrections and soft gluon corrections; b)
corrections containing an extra resolvable gluon or quark in the final
state (the dominant diagram is $gg\to ag$) in close proximity to the
$a$.  The total cross sections predicted by HIGLU~\cite{Spira:1996if}
are plotted as a function of $\ma$ for $\cabb=1/\catt=\tanb=1,2,3,10$
in Fig.~\ref{totsigs} with and without the resolvable parton final
state QCD corrections. The HIGLU results agree well with a private
program for this process.  We note that the cross sections do not
scale precisely as $\tan^2\beta$ at large $\tanb$ (as naively
predicted by dominance of the $b$-quark loop diagram for the $gg\to a$
coupling at high $\tanb$) due to the virtual corrections. In any case,
very substantial cross sections are predicted. In the NMSSM context,
at any given $\tanb$ value one should reduce the plotted result for
that $\tanb$ by a factor of $(\cta)^2$.

Among the cuts employed in the CDF analysis there is an isolation
requirement whereby events are only included if both muons have less
than $4\gev$ scalar summed $p_T$ in a cone of size $\Delta R=0.4$
about the muon.  The impact of the isolation requirement was studied
for the $\upsi$ and it was found that this isolation requirement was
99.8\% efficient for the $\upsi$ despite the fact that $\upsi$'s
are produced along with one or more extra particles in the final state.  Thus,
in our analysis for the $a$ we will make the assumption that the
components of the $a$ cross section coming from final states
containing an extra $q$ or $g$ are not significantly affected by the
isolation cut and thus we will employ the full QCD-corrected $a$ cross
section.  In addition, in the analysis of \cite{Aaltonen:2009rq} only
events for which the $\mupmum$ pair resides in the $|y|<1$ region are
retained.  Thus, what we actually employ are the cross sections
\beq
\sigma(a)_{|y|<1}=\left.{d\sigma(a)\over dy}\right|_{y=0}\times 2\,,
\eeq
which is an excellent approximation given that the cross section is
essentially flat in $y$ over this region.  At the Tevatron, the ratio 
\beq
{\left.{d\sigma(a)\over dy}\right|_{y=0}\over \sigma(a)_{tot}}
\eeq
varies from roughly $0.12$ at $\ma\sim 2\gev$ to $0.19$ at $\ma\sim 12\gev$
with very weak dependence on $\tanb$.  At $\ma=\mupsi$ the ratio is
$\sim 0.15$. 

In \cite{Aaltonen:2009rq}, what is given are limits on the ratio for production
of a very narrow resonance, the $\eps$, relative to
that  for the $\upsi$
\beq
R={\sigma(\eps)\br(\eps\to \mupmum)\over \sigma(\upsi)\br(\upsi\to\mupmum)}
\label{rdef}
\eeq
under the assumption that the same mechanism is responsible for $\eps$
production as is responsible for $\upsi$ production.  As stated
earlier, since the $a$ can be produced directly via $gg\to a$ whereas
the $\upsi$ cannot, an interpretation for the $a$ of the limits given
for the generic $\eps$ requires actually knowing what the
$\upsi$ cross section is. It also requires an assumption
regarding the efficiency for the $a$ of acceptance and isolation
requirements relative to those employed for the $\eps$.

Our analysis is the following. In Ref.~\cite{Apollinari:2005fy}, it is
stated that the cross section for $\upsi$ production in the
$|y|<0.6$ region at $\rts=1.8\tev$ was measured to be $34,600\pb$.  In
contrast, the cuts of Ref.~\cite{Apollinari:2005fy} accept $\upsi$
events with $|y|<1$. In Ref.~\cite{Apollinari:2005fy}, it is stated
that the efficiency for $\upsi$ detection (due to
geometric and kinematic acceptance cuts as well as trigger and
reconstruction efficiencies, but before imposing the isolation
requirement noted above) is $0.066$. From \cite{Abe:1995an}, we infer
that this acceptance times efficiency factor is one that applies at
any fixed value of $y$ that is relatively central and after
integrating over accepted $p_T$'s.
Then, using $\br(\upsi\to\mupmum)=0.0248$ and
the integrated luminosity of $L=110\pbi$ one then predicts
\beq 34600\times\left({2\over
    1.2}\right)\times 110\pbi\times 0.0248 \times 0.066=10383 
\label{estimate}
\eeq
events where the parenthetical fraction corrects for the increased $\dely$ acceptance 
compared to that used in measuring the $\upsi$
cross section.  This compares favorably to the 9838 number of events
that were observed before including 
the isolation cuts and promptness cuts of Table 1 in
Ref.~\cite{Apollinari:2005fy}.  A cross check on the cross section is to note that the
$\left.{d\sigma(\upsi)\over dy}\right|_{y=0}\times \br(\upsi\to \mupmum) \sim 753\pb$ value
measured in \cite{Abe:1995an} is comparable to the estimate based on
Eq.~(\ref{estimate}) of
\beq
\left.{d\sigma(\upsi)\over dy}\right|_{y=0}\times \brupsimumu
={\sigma(\upsi)_{|y|<0.6}=34600\pb \over \dely=1.2}\times 0.0248\sim 715\pb\,.
\label{xsecest}
\eeq 
Because the earlier paper \cite{Abe:1995an} may
have employed slightly different procedures, efficiencies and so
forth, we use the value of Eq.~(\ref{xsecest}) at $\sqrt s=1.8\tev$.

Moving to the higher energy of $\rts=1.96\tev$, it is stated in
\cite{Aaltonen:2009rq} that the $|y|<0.6$ $\upsi$ cross section
increases relative to $\rts=1.8\tev$ by about 10\%, implying
\beq
\left.{d\sigma(\upsi) \over dy}\right|_{y=0}(1.96\tev)\times\br(\upsi\to
\mupmum) \sim 787 \pb\,.
\label{dsigdybr1.96}
\eeq 
This is the value we shall employ. As another cross check, we note
that a 10\% increase in the total cross section would yield
about $38,330\pb$ at $\rts=1.96\tev$.  
At $\rts=1.96\tev$, \cite{Aaltonen:2009rq} states that
52,700 $\upsi\to\mupmum$ events are observed using the same
cuts as in Ref.~\cite{Apollinari:2005fy} (that imply that only an
$\upsi$ with $|y|<1$ will be accepted) {\it and} after imposing the
isolation and promptness criteria detailed in
Ref.~\cite{Apollinari:2005fy}. The latter imply an additional efficiency
factor of $0.921$ relative to the $0.066$ efficiency referenced
earlier. Multiplying these two efficiencies yields a net efficiency of
$\sim 0.061$. With the 10\% cross section increase and
accounting for the increased luminosity of $L=630\pbi$, the $0.061$
net efficiency implies an
expected event number of 60,244. Although this is not in perfect agreement
with the 52,700 events actually observed, we will use the
result of Eq.~(\ref{dsigdybr1.96}) below.

Relative to the $\upsi$
efficiency, purely geometric effects alter the efficiency for $\eps$
production and we assume that the same geometric changes apply to the
$a$. The formula of \cite{Apollinari:2005fy} is:
\beq
efficiency(\eps) = efficiency(\upsi)\left[0.655+{(0.974-0.655)\over
    (9.0-6.3)} (m_\eps-6.3)\right]
\label{efficiency}
\eeq
We will employ this same relative efficiency for the $a$ as a function of
$\ma$ using as well $efficiency(\upsi)=0.061$ as obtained above.

The most precise limits on $\cabb$ are obtained using the ratio $R$
defined in Eq.~(\ref{rdef}).  We recall from
Refs.~\cite{Apollinari:2005fy,Aaltonen:2009rq} that the limits on $R$
are obtained by performing a smooth fit to the event distribution and
looking for fluctuations about this smooth fit.  The limits on the
$\eps$ (or the $a$ in our case) are then obtained by placing small
Gaussians at each possible $\ma$ value and placing limits using the
observed fluctuations about the smooth fit. In the mass region for
which CDF has performed this analysis, $6.3\gev\lsim \ma\lsim 9\gev$,
it is very convenient to simply directly employ their results.  As
stated, we assume that the $a$ efficiencies are the same as for the
$\eps$, in which case we can compute the ratio $R$ as
\beq
R\simeq {\left.{d\sigma(a)\over dy}\right|_{y=0}\times \br(a\to \mupmum)\over
  \left.{d\sigma(\upsi)\over dy}\right|_{y=0}\times \brupsimumu}
\label{rcomp}
\eeq
where $\left.{d\sigma(a)\over dy}\right|_{y=0}$ is computed using
HIGLU, $\br(a\to \mupmum)$ is taken from Fig.~\ref{bramumu} and
$\left.{d\sigma(\upsi)\over dy}\right|_{y=0}\brupsimumu$ is as given
in Eq.~(\ref{dsigdybr1.96}).  Note that the exact values of the
efficiencies, Eq.~(\ref{efficiency}), are not important using this
procedure so long as the efficiency for the $a$ is the same as for the
$\eps$.

\begin{figure}[h!!]
\begin{center}
\includegraphics[width=0.62\textwidth,angle=90]{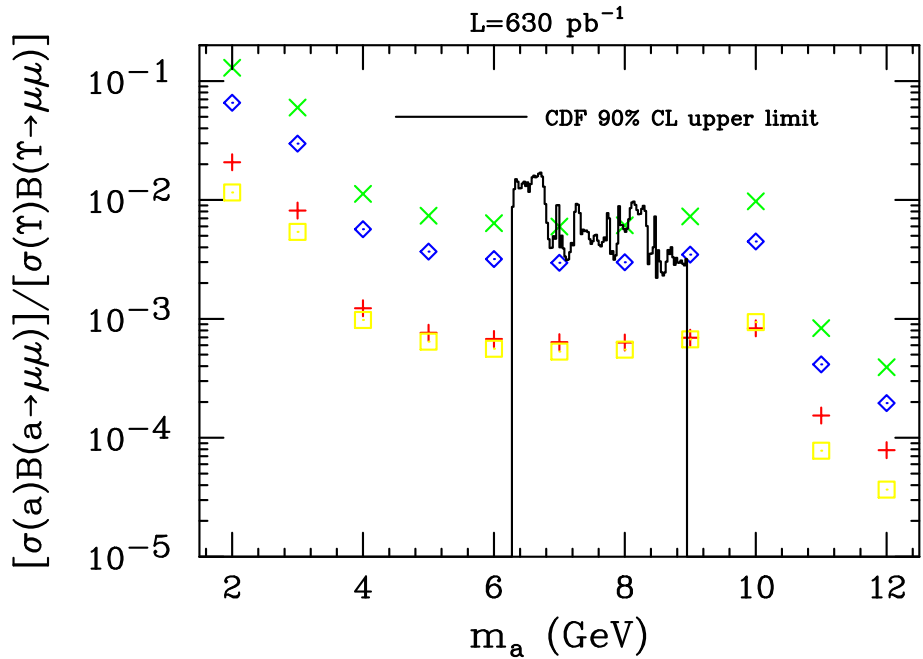}
\end{center}
\vspace*{-.3in}
\caption{We plot the 90\% CL limits on the ratio $R$ for $a$
  production at the Tevatron as a function of $\ma$ compared to NMSSM
  predictions using HIGLU for the following cases: $(\tanb=1,\cta=1)$
  (red $+$'s), $(\tanb=2,\cta=1)$ (blue diamonds), $(\tanb=3,\cta=1)$
  (green $\times$'s) and $(\tanb=10,\cta=0.1)$ (yellow squares).  }
\label{90percentCL}
\vspace*{-.05in}
\end{figure}

\begin{figure}[h!]
\begin{center}
\includegraphics[width=0.62\textwidth,angle=90]{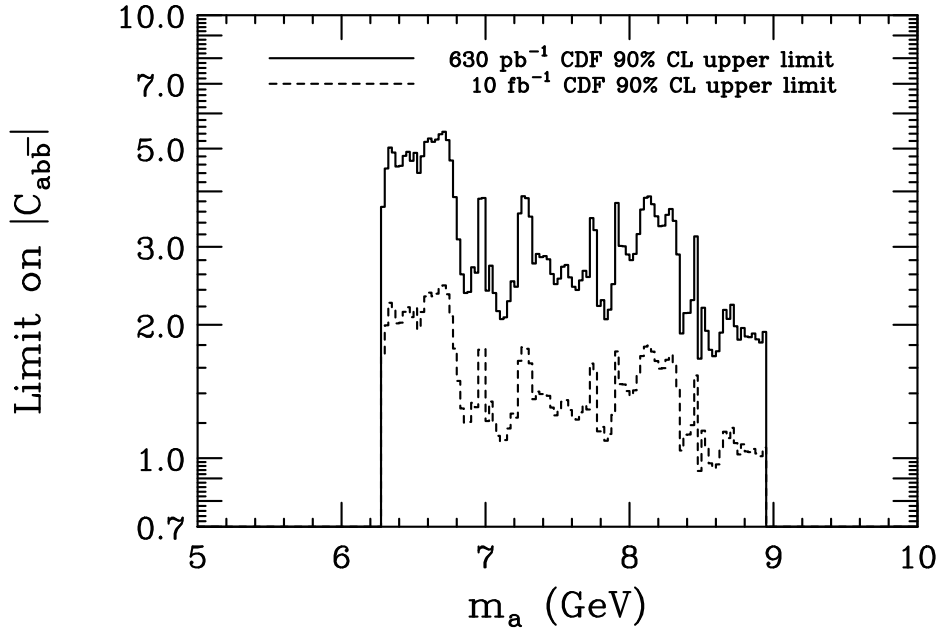}
\end{center}
\vspace*{-.3in}
\caption{We plot the $90\%$ CL upper limits on $|\cabb|$ obtained
  using the results for the ratio $R$ of
  Fig.~\protect\ref{90percentCL}. The $10\fbi$ results are obtained by
  statistical extrapolation of the $630\fbi$ results. In the context
  of the 2HDM(II), $\cabb=\tanb$.  In the context of the NMSSM,
  $\cabb=\cta\tanb$. In both cases, limits were derived
  assuming $\tanb\geq 1$. }
\label{90percentCL2lums}
\end{figure}

With these assumptions and inputs we can then predict the ratio $R$
for the case of $\eps=a$ and compare to the 90\% CL upper limits of
\cite{Aaltonen:2009rq} based on $L=630\pbi$ of analyzed CDF data.
This comparison appears in Fig.~\ref{90percentCL} for a number of
$\cabb=\cta\tanb$ choices. We observe that the predicted $R$ depends
almost entirely on $|\cabb|$, with extremely little dependence on
$\tanb$ separately for the $\tanb\geq 1$, $\ma\geq 4\gev$ parameter
region on which we focus. The corresponding bin-by-bin limits on
$|\cabb|$ obtained by interpolation appear in
Fig.~\ref{90percentCL2lums}.  In the 2HDM(II), they are limits on
$\tanb=\cabb$.  In the NMSSM, these are limits on
$\cabb=\cta\tanb$. In both cases, the interpolations are only accurate
for $\tanb\geq1$ and $\ma\geq 4\gev$. From
Fig.~\ref{90percentCL2lums}, we find that the limits based on the
existing $L=630\pbi$ analysis roughly exclude $|\cabb|>3$ for
$6.8\lsim\ma\leq 9\gev$ and $|\cabb|>2$ for $8.2\lsim \ma\leq 9\gev$,
but do not exclude $\cabb=1$ for any of the $\ma$ values in the analysis range.
In the 2HDM(II) case the $\cabb=\tanb$ limits from the Tevatron are
stronger than those from Upsilon decays and LEP data, as summarized
earlier, for $8\gev\lsim \ma \lsim 9\gev$. In the NMSSM case the
limits on $|\cta|$ from the Tevatron data are the stronger in much the
same mass range, as we detail shortly. In Fig.~\ref{90percentCL2lums}
we also plot the statistically extrapolated limits that would result
by increasing the data sample to $L=10\fbi$. (Presumably, a real
analysis of a high luminosity data set would do even better.)  Since
the $a$ signal cross section varies roughly as $(\cabb)^2$, even this
large luminosity increase leads to limits that are improved by only a
factor of a bit more than two.  Nonetheless, one approaches the
$|\cabb|\sim 1$ level of interest in the NMSSM at $\ma=9\gev$.

\begin{figure}
\begin{center}
\includegraphics[width=0.65\textwidth,angle=90]{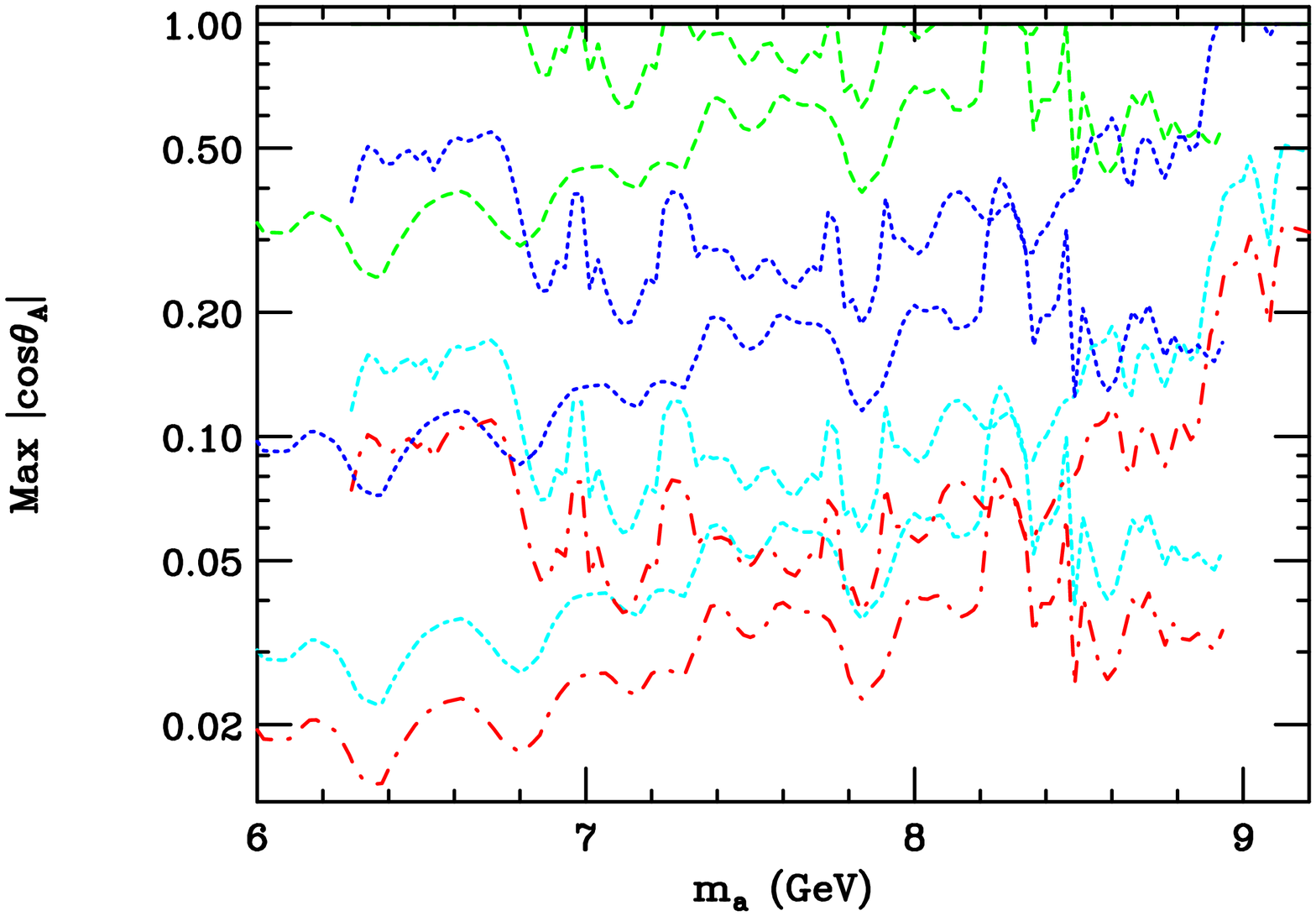}
\end{center}
\caption{We plot the $90\%$ CL $\ctamax$ values in the NMSSM
  context obtained from the results of Fig.~\protect\ref{90percentCL}
  for the $630\pbi$ CDF data set, in comparison to the $\ctamax$ values
  plotted in Fig.~\protect\ref{ctamaxvsma}. For clarity, the plot is limited
  to the $\ma$ region over which the Tevatron data are relevant.
The curve types are as in Fig.~\protect\ref{ctamaxvsma}. The
Fig.~\protect\ref{ctamaxvsma} results for a given curve type are those for
which $\ctamax$ starts at lower values at low $\ma$ rising to higher
values at higher $\ma$. The new CDF limits are those that begin near
$\ma\sim 6.3\gev$ and terminate at $\ma\sim 9\gev$ and that fall (with
fluctuations) as $\ma$ increases.}
\label{ctamaxtevatron}
\end{figure}

Focusing now on the NMSSM, we compute the upper limit on $\cta$,
$\ctamax$, obtained by appropriate interpolation of the results of
Fig.~\ref{90percentCL}.  We again emphasize that although $\br(a\to
\mupmum)$ is $\tanb$ dependent as shown in Fig.~\ref{bramumu}, it is
nonetheless the case that for $\tanb\geq 1$ and $\ma\geq 4\gev$ the
limits on $\cabb=\cta\tanb$ are almost independent of $\tanb$ at fixed
$\cabb$, as found in Fig.~\ref{90percentCL} (compare the two
$\cta\tanb=1$ cases --- $\tanb=\cta=1$ vs. $\tanb=10$,
$\cta=0.1$). This can be understood as follows. At low $\tanb$,
although $\br(a\to\mupmum)$ is suppressed, contributions to the $gga$
coupling from loops involving the top quark are substantial relative
to loops involving the bottom quark. In comparison, at large $\tanb$
one finds that $\br(a\to\mupmum)$ is maximal but top-quark loops are
relatively suppressed compared to bottom quark loops.  These two
effects very nearly cancel one another leaving the net $a$ cross
section unchanged at fixed $\cabb=\cta\tanb$.  As a result, it is easy
to extract the $\ctamax$ values for different $\tanb$ values directly
from the plotted limit on $|\cabb|$ shown in
Fig.~\ref{90percentCL2lums}.

The resulting $\ctamax$ limits are shown in Fig.~\ref{ctamaxtevatron}
in comparison to the upper limits plotted earlier in
Fig.~\ref{ctamaxvsma}. The figure focuses on the $6\gev\lsim\ma\lsim
9\gev$ region for which we have extracted the Tevatron limits using
$R$.  What we observe is that the $630\pbi$ 90\% CL limits become the
strongest for $\ma\gsim 8.3\gev$. In a forthcoming paper, we will
analyze the impact of BaBar data for $\upsiii\to\gam \tauptaum$ decays
on the $6\gev\lsim\ma\lsim 10\gev$ region.  Our preliminary results
suggest that the Tevatron limits plotted above and the $\upsiii$
limits are very similar at $\ma\sim 9\gev$, with $\upsiii$ limits
being superior for lower $\ma$. 

Given the above, the great value of extending the Tevatron analysis
above $\mmumu=9\gev$ is apparent. A full analysis of existing and
future data all the way out to $\ma\sim 12\gev$ is needed. First, it
might strongly constrain the properties of any light $a$ with
$\ma\lsim 2m_B$ that would allow for the ideal Higgs scenario. Second,
it might completely eliminate the possibility that a light $a$ could
provide a major contribution to $a_\mu$. At the moment, the Tevatron
limits on $\cabb$ shown rule out a significant contribution to $\Delta
a_\mu$ from an $a$ with $\ma<9\gev$, while in the range
$9\gev\lsim\ma\lsim 2m_B$ these limits leave open the possibility of
$\Delta a_\mu$ arising from diagrams involving a CP-odd $a$. Only the
Tevatron and/or LHC can probe the region of $\ma$ above the Upsilon
masses.

\begin{figure}[h!]
\begin{center}
\includegraphics[width=0.65\textwidth,angle=90]{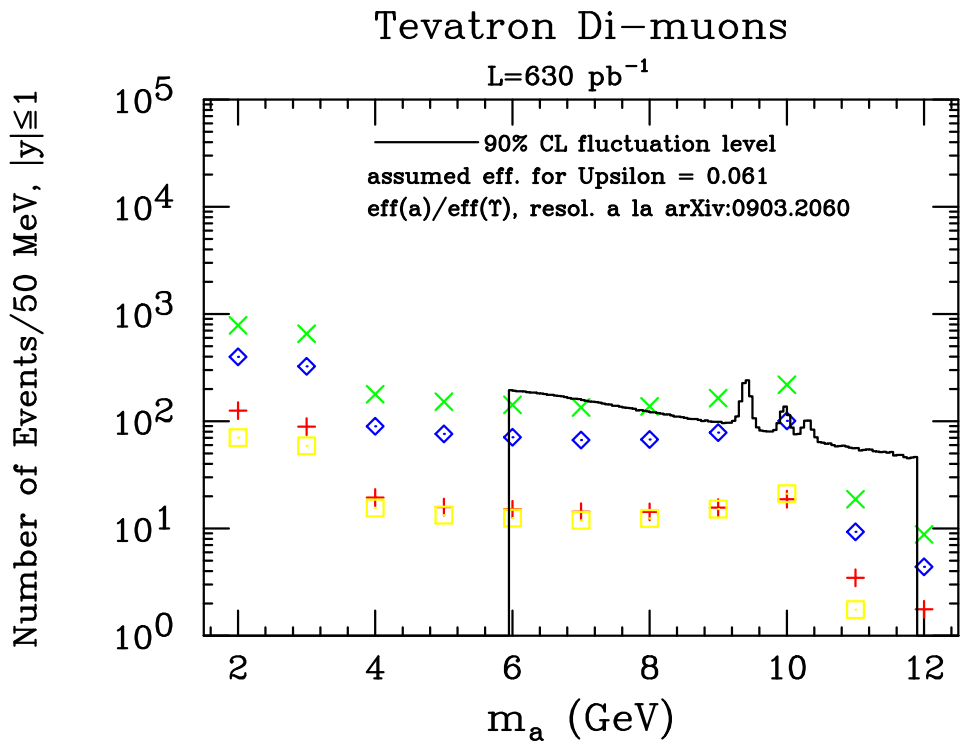}
\end{center}
\caption{We plot the upwards fluctuation in the number of events in a given bin
  corresponding to a $1.646\sigma$ (90\% CL) excess as predicted using the
  $L=630\pbi$ event numbers of \cite{Aaltonen:2009rq}.  These limits
  are compared to the predicted number of events for an $a$ resonance
  centered on that bin spread out by the experimental resolution.  The
  same $(\tanb,\cta)$ values as in Fig.~\protect\ref{90percentCL} are considered.  }
\label{90percentCLnevt630pbi}
\end{figure}

Absent a full analysis by CDF of limits on $R$ in the region
$\mmumu>9\gev$ and given that this region is of great interest, we
wish to make some estimates of limits on $|\cabb|$ based on the event
number plots of Ref.~\cite{Aaltonen:2009rq}.  We have employed the
following procedure.  First, for this analysis, we must know the
efficiency for detecting the $a$. For our estimates we use $efficiency(\ma)$
from Eq.~(\ref{efficiency}) and $efficiency(\upsi)=0.061$ (as
motivated earlier below Eq.~(\ref{dsigdybr1.96})) to predict the
number of $a$ events as a function of $\ma$.  Second, we wish to
determine how many of the total number of $a$ events fall into a
$50\mev$ bin centered on $\ma$. To do so, we need to know the
resolution as a function of $\ma$. In~\cite{Aaltonen:2009rq}, it is
stated that the resolution, $\sigma_r$ varies from $32\mev$ to
$50\mev$ in going from $\ma=6.3\gev$ to $\ma=9\gev$, with a value of
$52\mev$ at $\mupsi$. We use a simple linear interpolation for other
values of $\ma$, but do not allow $\sigma_r$ to fall below $25\mev$ at
low $\ma$. The fraction of $a$ events distributed as a Gaussian
of width $\sigma_r$ that fall into a $50\mev$ bin (which should be thought of
as a bin of half width $25\mev$) that is centered on $\ma$ is given by 
\beq
f(m)=Erf\left({25 \mev\over \sqrt{2} \sigma_r(m)}\right)
\eeq
where $\sigma_r(m)$ is in $\mev$.
In Fig.~\ref{90percentCLnevt630pbi}, we plot the $1.646\sigma$, \ie\
90\% CL, fluctuation number for each of the CDF $50\mev$ bins compared
to the predicted number of $a$ events that would fall into that bin.
We do this for the same selection of $(\tanb,\cta)$ values as employed
in Fig.~\ref{90percentCL}. One observes that for $\ma\sim 6\gev$
($\ma\sim 9\gev$) 90\% CL sensitivity is anticipated for $\cabb\gsim
3$ ($\cabb\gsim 2$). This anticipates in an average sense the more
precise (and more fluctuating) results based on the $R$ analysis found
in Fig.~\ref{90percentCL2lums}.

\begin{figure}
\begin{center}
\includegraphics[width=0.65\textwidth,angle=90]{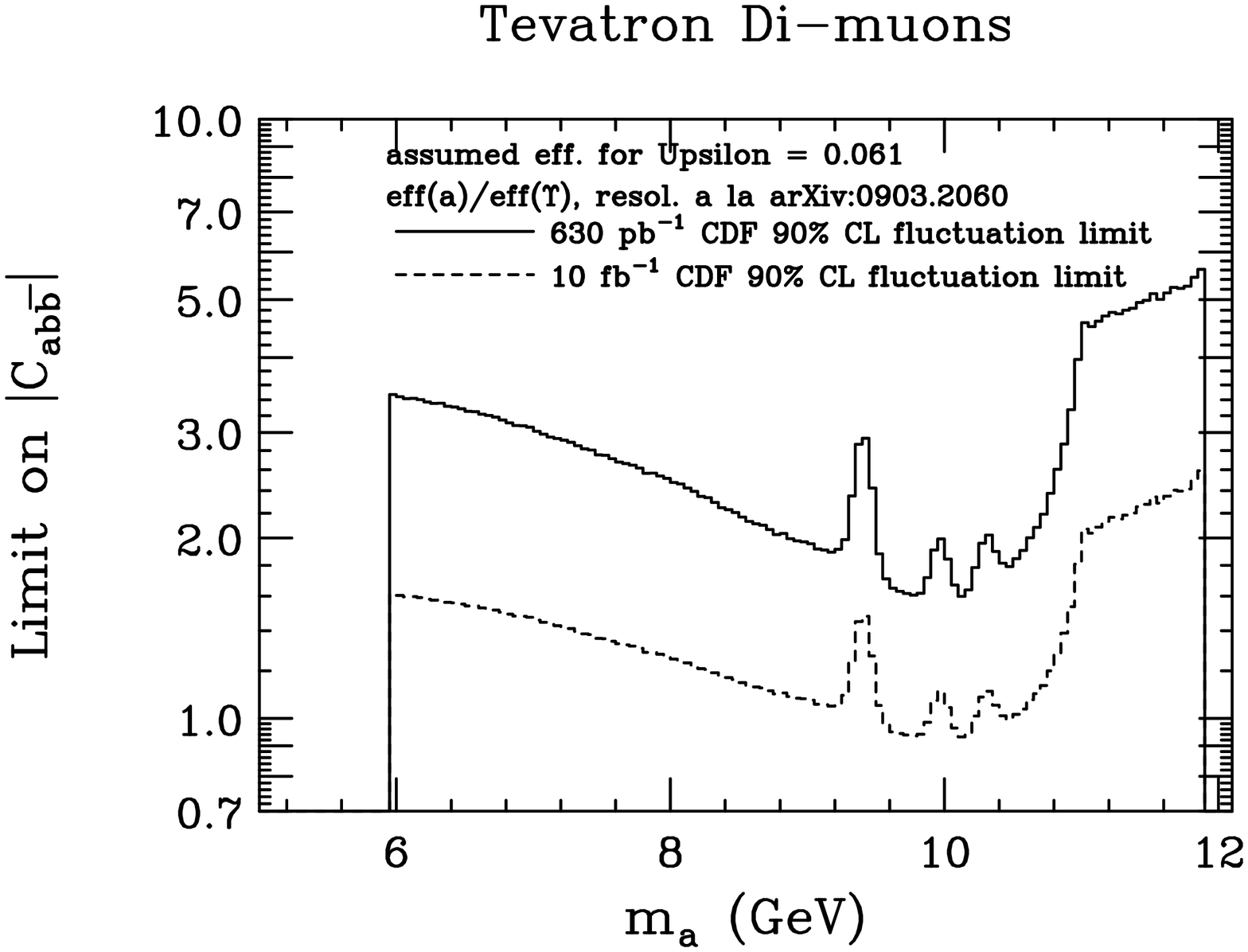}
\end{center}
\caption{We plot approximate limits on $|\cabb|$ as a function of
  $\ma$ estimated assuming that bins centered on $\ma$ and
  encompassing an $\ma$ range of $2\sqrt{2}\times \sigma_r$, where
  $\sigma_r$ is the experimental $\mmumu$ resolution at the given
  $\ma$, do not have a $90\%$ CL fluctuation relative to the number of
  events observed and plotted over this bin range in
  \cite{Aaltonen:2009rq} for $L=630\pbi$.  The $L=10\fbi$ histogram
  corresponds to simply scaling the predicted $a$ event rate and the
  $90\%$ CL fluctuations to the increased luminosity.  }
\label{tanblim_tevatron_2lums_evts}
\end{figure}

Sensitivity to the $a$ in the $S/\sqrt B$ sense can actually be
improved by taking a bin size that properly matches $\sigma_r$.  If
the background is flat then the optimal bin size is $2\sqrt 2
\sigma_r$ which retains a fraction $Erf(1)=0.843$ of the total $a$
signal and yields $B=2\sqrt 2 \sigma_r {d \sigma_B\over d\mmumu}$.
Following this procedure we can then use interpolation to extract the
$|\cabb|$ value such that $S/\sqrt B$ is $1.646$.  The resulting
values of $|\cabb|$ which correspond to this 90\ CL fluctuation in the
$\Delta \mmumu=2\sqrt 2 \sigma_r$ acceptance window are plotted in
Fig.~\ref{tanblim_tevatron_2lums_evts}.  As anticipated above, this
event counting method turns out to give a good average representation
of the results obtained using $R$ (which analysis was based on
bin-by-bin fits of the fluctuations about a smooth curve) at the 90\%
CL. Thus, despite relatively small $S/B$ levels (typically of order
0.02 in each of two neighboring bins for a $1.646\sigma$ net
fluctuation), our estimates for expectations for $\ma>9\gev$ (using
the approach of assuming there were no $1.646\sigma$ fluctuations in
the absolute $L=630\pbi$ event numbers in acceptance windows of size
$\Delta \mmumu$) should give a good idea of the limits that are
implicit in current data.

As an aside, we note that even though the shape of the $\Upsilon(nS)$
resonances (which are also very narrow) will also be determined by
$\sigma_r$, one can learn if there is an excess in the $\mupmum$ final
state by also looking at the $\Upsilon(nS)$ resonance in the $\epem$
final state to which the $a$ will not contribute.  Assuming lepton
universality, the $\Upsilon(nS)$ contribution in the $\mupmum$ final state
can be subtracted from the $\mupmum$ spectrum, after which any
residual excess from the presence of an $a$ would become
apparent. Statistical errors resulting from this subtraction will be
roughly a factor of $\sqrt 2$ larger than employed above. However,
this procedure does rely on a precise understanding of 
efficiencies, resolutions and the like for electrons. Another
technique that could be considered is comparing one Upsilon resonance
to another. If the relative normalization between the two Upsilon
resonances can be sufficiently precisely predicted, including both
theoretical and experimental uncertainties, then an $a\to \mupmum$
signal hiding under one of the Upsilons could become apparent.

\begin{figure}[h!]
\begin{center}
\includegraphics[width=0.65\textwidth,angle=90]{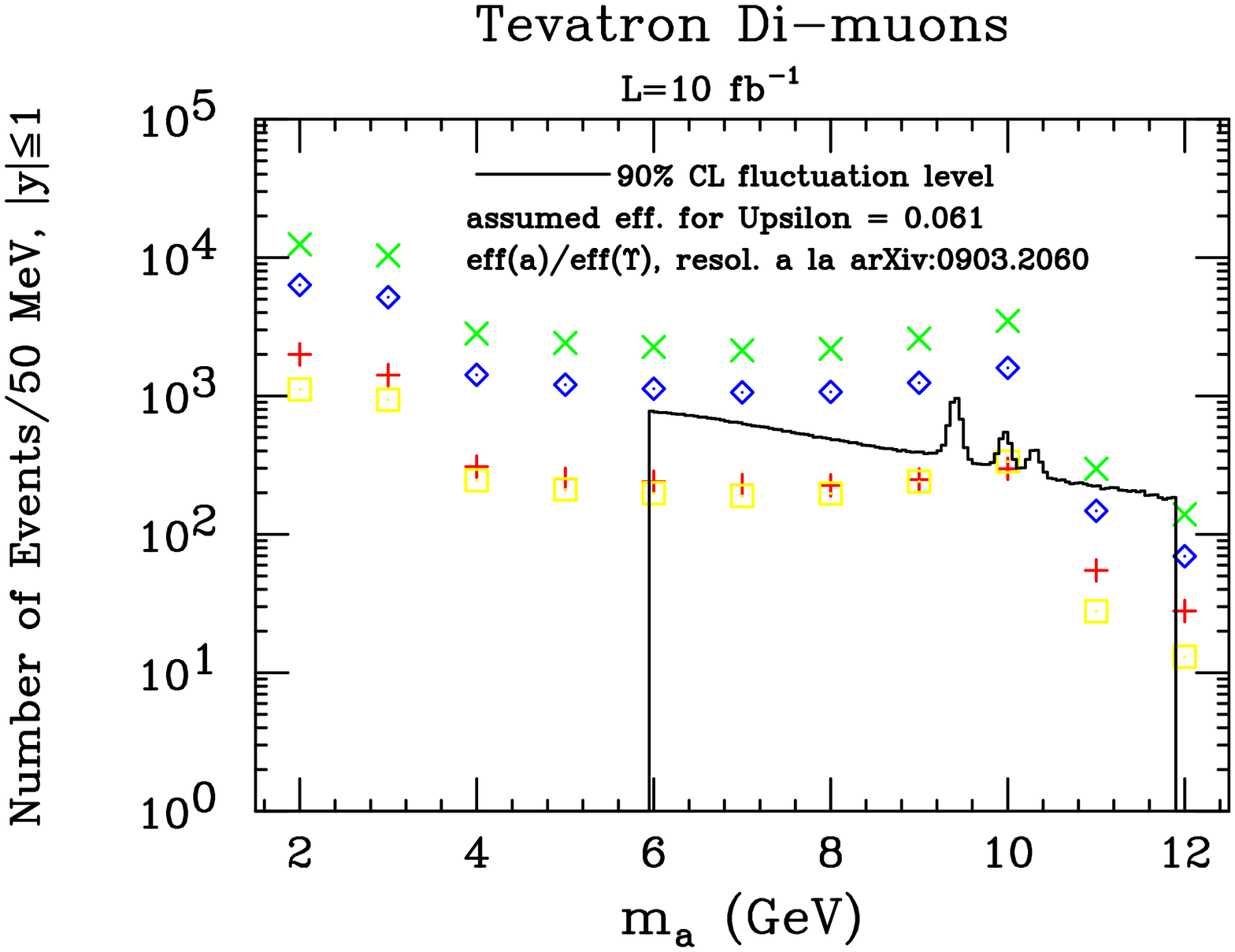}
\end{center}
\caption{We plot the upwards fluctuation in the number of events in a
  given bin corresponding to a $1.646\sigma$ (90\% CL) excess as
  predicted using the $L=630\pbi$ event numbers of
  \cite{Aaltonen:2009rq} scaled up to $L=10\fbi$.  These limits are
  compared to the predicted number of events for $L=10\fbi$ for an $a$ resonance
  centered on that bin spread out by the experimental resolution.  The
  same $(\tanb,\cta)$ values as in Fig.~\protect\ref{90percentCL} are
  considered.  }
\label{90percentCLnevt10fbi}
\end{figure}

Both CDF and D0 will continue to accumulate data far in excess of
$L=630\pbi$. Thus, it is useful to extrapolate to higher luminosity
using the observed number of events for $L=630\pbi$ plotted in
\cite{Aaltonen:2009rq}. We rescale the observed number of events in
each bin to $L=10\fbi$ and compute the $90\%$ CL fluctuation upper
limit in each bin as $1.646\times\sqrt{ N_{evt}(bin)}$.  In
Fig.~\ref{90percentCLnevt10fbi}, we plot these extrapolated $1.646\sigma$
fluctuation levels in each bin compared to the predicted number of $a$
events in each bin for the same selection of $(\tanb,\cta)$ values as
employed in Fig.~\ref{90percentCL}.  The extracted limits on $|\cabb|$
are plotted in Fig.~\ref{tanblim_tevatron_2lums_evts}.  We see that
the eventual Tevatron limits from just one experiment could easily be
superior to those currently available from Upsilon decays, in
particular probing the $\cabb=\cta\tanb\lsim 1$ coupling level, of
particular interest for the ``ideal'' Higgs scenario, all the way up to
$\ma=2m_B$, except in the vicinity of the Upsilon peaks.

Some further comments are the following.
First, we emphasize that the $1.646\sigma$ approximate procedure leads to the
expectation that quite important limits are possible for $L=10\fbi$ in
the $9\gev\lsim \ma\lsim 2m_B$ region of great interest in the NMSSM
version of the ideal Higgs scenario for which $\ma\gsim 8$ and
$|\cabb|\sim 0.2 - 2$ is a strongly preferred parameter region.  If this
scenario were to be nature's choice, there is a decent chance of
observing an $a$ using the dimuon spectrum analysis and the ultimate
Tevatron data set. 

Second, we again note that even the $L=630\pbi$ estimated limits of
Fig.~\ref{tanblim_tevatron_2lums_evts} for $9\gev\leq \ma\leq 12\gev$
would rule out the enhanced $|\cabb|$ values of order $30$ needed for
$a$-exchange graphs to explain the $a_\mu$ discrepancy for $\ma$ in
this mass region, which is the only relatively low mass region for
which other current constraints are sufficiently weak that the $a$
might provide the observed discrepancy.  It is thus quite important
for CDF (and D0) to perform the needed analysis using the $R$ or
similar technique and determine whether or not the rough limits we
obtained above using event numbers are approximately correct.

\section{LHC Prospects}
\label{lhc}

\begin{figure}
\begin{center}
\includegraphics[width=0.65\textwidth,angle=90]{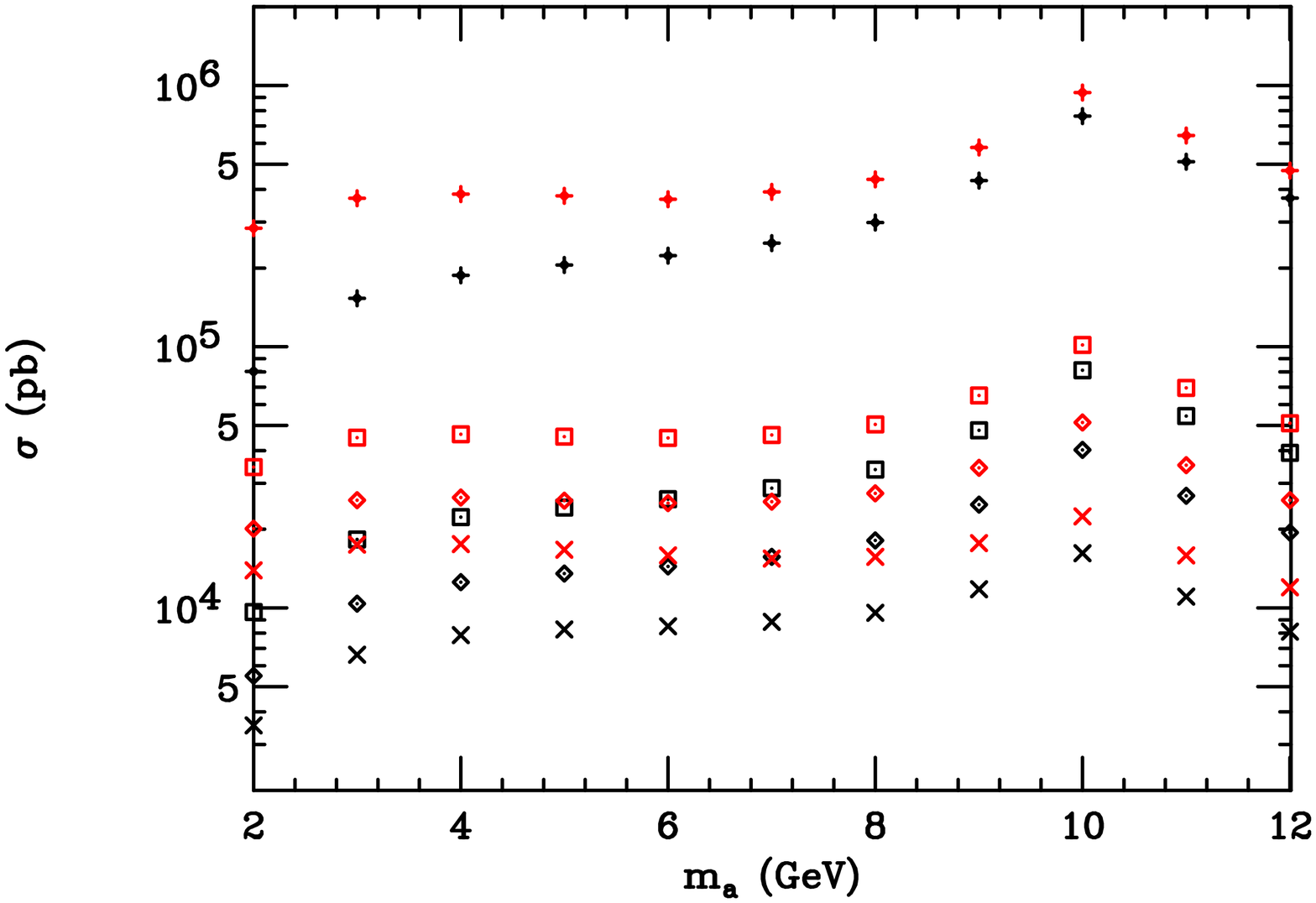}
\end{center}
\caption{The total cross section for $a$ production at the LHC
  for $\rts=14\tev$ is plotted vs $\ma$ for $\tanb=1,2,3,10$ (lowest to
  highest point sets).  For each $\ma$ and $\tanb$ value, the lower
  (higher) point is the cross section without (with) resolvable parton
  final state contributions.  }
\label{totsigslhc}
\end{figure}

\begin{figure}
\begin{center}
\includegraphics[width=0.65\textwidth,angle=90]{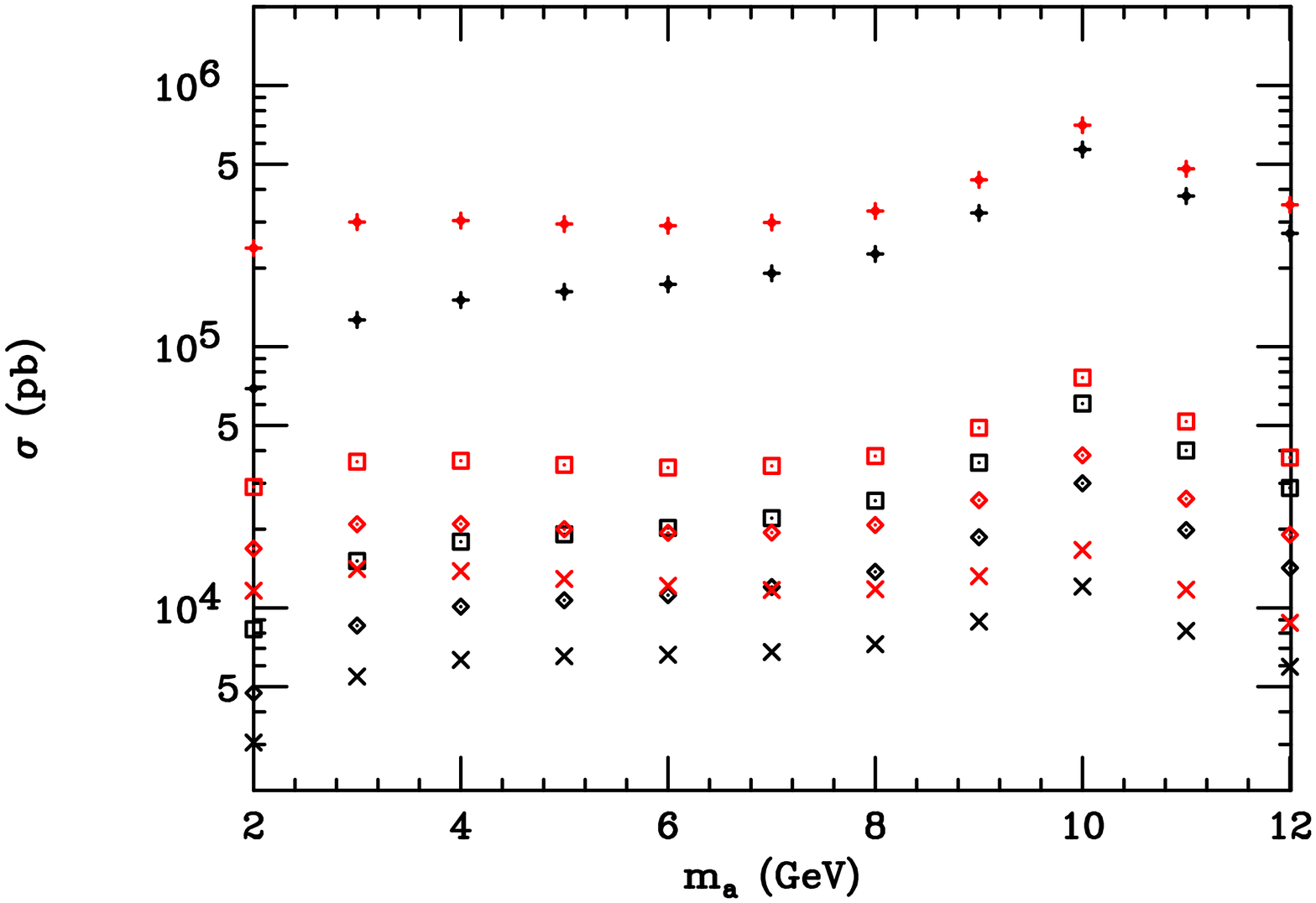}
\end{center}
\caption{The total cross section for $a$ production at the LHC
  for $\rts=10\tev$ is plotted vs $\ma$ for $\tanb=1,2,3,10$ (lowest to
  highest point sets).  For each $\ma$ and $\tanb$ value, the lower
  (higher) point is the cross section without (with) resolvable parton
  final state contributions.  }
\label{totsigslhc10}
\end{figure}

\begin{figure}
\begin{center}
\includegraphics[width=0.65\textwidth,angle=90]{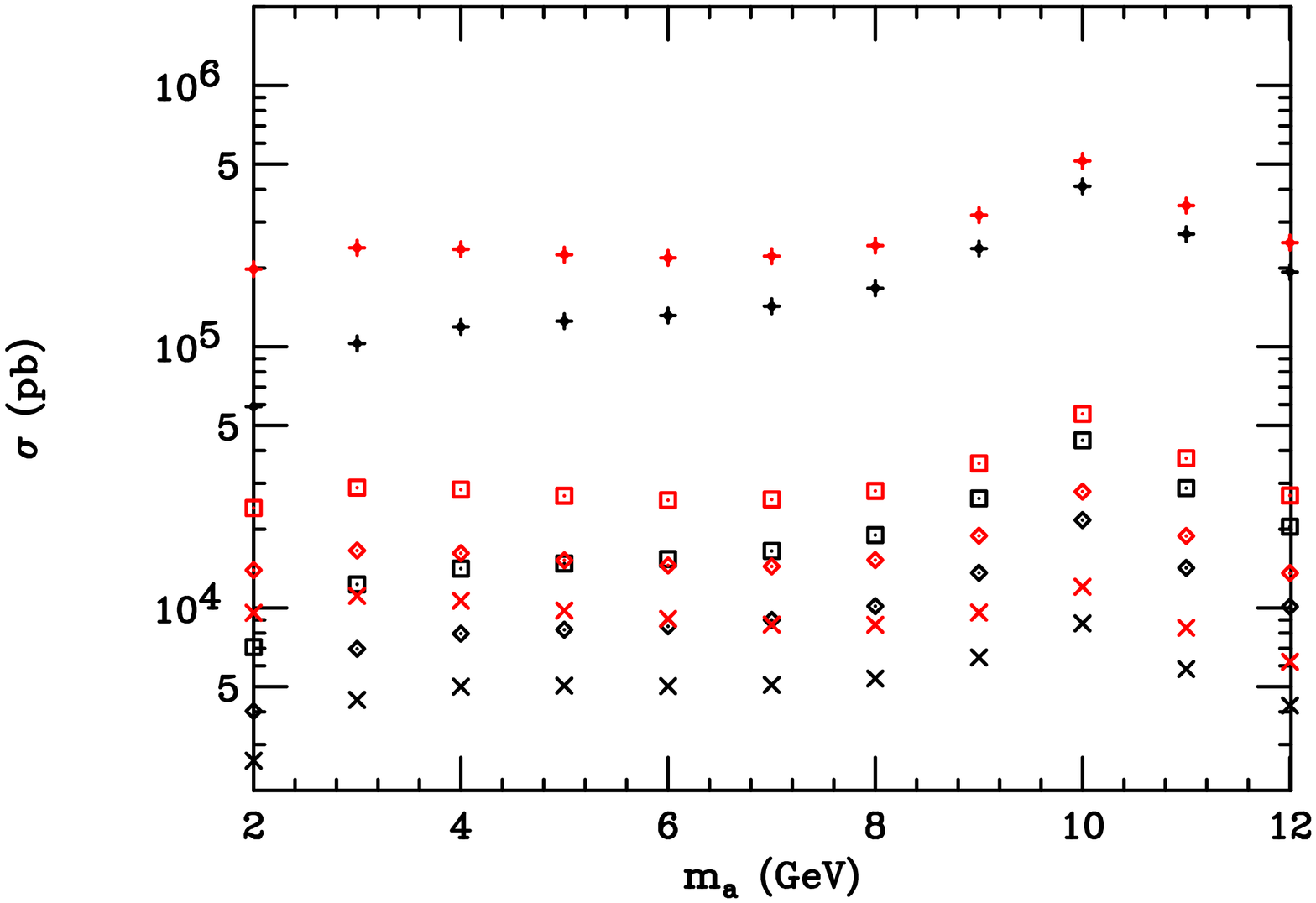}
\end{center}
\caption{The total cross section for $a$ production at the LHC
  for $\rts=7\tev$ is plotted vs $\ma$ for $\tanb=1,2,3,10$ (lowest to
  highest point sets).  For each $\ma$ and $\tanb$ value, the lower
  (higher) point is the cross section without (with) resolvable parton
  final state contributions.  }
\label{totsigslhc7}
\end{figure}

The basic question is whether the LHC will be able to improve over the
Tevatron $L=10\fbi$ projected results and limits.  The total cross
sections at the LHC appear in Figs.~\ref{totsigslhc} and
\ref{totsigslhc10} for $\rts=14\tev$ and $\rts=10\tev$ respectively;
they are plotted analogously to those for the Tevatron appearing in
Fig.~\ref{totsigs}. Recall that these cross sections are those
appropriate in the 2HDM(II) context. We see that, relative to the
Tevatron, the $\rts=14\tev$ cross sections are about a factor of $3-7$
higher, the smaller (larger) ratio applying at small (large) $\ma$.
Relative to the $\rts=14\tev$ cross section, the $\rts=10\tev$ cross
sections are roughly a factor of $\sim 1.2$ smaller at $\ma=2\gev$ and
a factor $\sim 1.34$ smaller at $\ma=10\gev$, more or less independent
of the $\tanb$ value. It now appears that perhaps as much as a year
will be spent running at $\rts=7\tev$. Thus, we also plot in
Fig.~\ref{totsigslhc7} the cross sections for this latter energy.  At
$\ma=2\gev$, the $a$ cross section at $\rts=10\tev$ is a factor of
about 1.15 larger than the $\rts=7\tev$ cross section; the factor
rises to $\sim 1.35$ for $\ma=10\gev$. The modest decrease of the $a$
cross section with decreasing energy is a result of the fact that the
$gg$ luminosity at low $\ma$ varies slowly with $\rts$.  This is one
of the reasons why searches for a light $a$ are very appropriate in
early LHC running.

In going to the NMSSM, one takes these results
for any given $\tanb$ choice and then multiplies by $(\cta)^2$, the
square of the overlap fraction of the $a$ with the 2HDM component.

As an example of how limits obtained at the LHC in early running will
compare to the Tevatron limits, let us consider the case of $\tanb=10$
and $\cta=0.1$, for which $\cabb=1$.  As shown in
Fig.~\ref{tanblim_tevatron_2lums_evts}, even with $L=10\fbi$ the
Tevatron is not fully able to probe at the $90\%$ CL the predicted
relatively small $a$ event levels except at $\ma$ values close to
$2m_B$ but outside the $\Upsilon(nS)$ peaks.  In more detail, for the
above parameter choices, the predicted number of $a\to \mupmum$ events
for $|y|\leq 1$ in a $\Delta \mmumu=2\sqrt 2 \sigma_r$ bin centered on
$\ma$ is 436, 615 and 475 at $\ma=8\gev$, $\mupsi$ and $10.5\gev$,
respectively, where the event numbers quoted incorporate the
$Erf(1)=0.8427$ reduction factor associated with keeping only events
in an interval of size $\Delta \mmumu=2\sqrt 2 \sigma_r$. The actual
$\Delta \mmumu=2\sqrt{2}\sigma_r$ values are $43\mev$, $52\mev$ and
$57\mev$ at $8\gev$, $\mupsi$ and $10.5\gev$, respectively. As regards
the background, we take the
50 MeV bin event numbers in the CDF plot of the number of events in
each bin and rescale to the  $\Delta \mmumu$ interval sizes at the
above $\mmumu=\ma$ choices.  This gives us a background event
number $N_{\Delta \mmumu}$ at each $\ma$. The $1\sigma$ fluctuations in these
background event numbers, $\sqrt{N_{\Delta \mmumu}}$, are 468, 945 and
285, respectively.  The statistical significances of the $a$ signals
are then $\sim 0.93\sigma$, $\sim 0.65\sigma$ and $\sim 1.67\sigma$,
respectively. Only the latter is (slightly) above the $1.646\sigma$
level corresponding to 90\% CL.  However, to repeat, this high
$\ma\lsim 2m_B$ region is particularly favored in the model context.
But, to reach $5\sigma$ at $\ma=10.5\gev$ would require about nine
times as much integrated luminosity, \ie\ $L\sim 90\fbi$ and $5\sigma$
at $\ma=\mupsi$ would require $L\sim 590\fbi$.

Projections for the LHC have been made public by ATLAS.  In Fig.~1
of~\cite{priceatlas}, one finds a plot of $d\sigma/d\mmumu$ coming
from $b\anti b$ production, Drell-Yan production and $\upsi$
production. The dimuon Drell-Yan contribution is negligible compared to
that from $b\anti b$ production even after the latter is reduced by
muon isolation requirements. We ignore the Drell-Yan contribution in
all subsequent discussions. 

In generating the $b\anti b$ and $\upsi$ cross sections, only events
with $p_T$ cuts requiring one muon with $p_T>6\gev$ and a 2nd muon
with $p_T>4\gev$, both with $|\eta|<2.4$, were retained. A recent
Monte Carlo study \cite{student} finds that these events constitute
20\% of the total inclusive cross section. The fraction of these
events that survive after further requirements related to triggering,
reconstruction and the final analysis selection cuts is 50\%.  Thus,
the net efficiency for the $\upsi$ events plotted in Fig.~1 of
\cite{priceatlas} is $\sim 0.5\times 0.2=0.1$.  Therefore, we will
write $\effatlas=0.1 r$ for the fraction of inclusive $a$ events that
will be retained, where $r\sim 1$ for the cuts and triggering
strategies studied so far, but $r>1$ is probably achievable if these
are optimized for the CP-odd $a$.

Returning to Fig.~1 of \cite{priceatlas}, we observe a $b\anti
b$-induced dimuon cross section level for $\rts=14\tev$ of order
$d\sigma/d\mmumu\sim 50-90\pb/100\mev$ in the $\mmumu\in[8\gev,2m_B]$
interval when outside the Upsilon peak region.  This is the dimuon
cross section from $b\anti b$ heavy flavor production only.  The
author of ~\cite{priceatlas} estimates~\cite{dpprivatecom} that one
should at most double this cross section to account for $c\anti c$
production and other contributions.  We will make estimates based on
multiplying the $b\anti b$-induced dimuon cross section by a factor of
two.  To this, we add the $\upsi$ cross section as plotted in Fig.~1.
The net resulting spectrum constitutes the background to the $a$
signal that we discuss shortly.

As in the CDF case, we will use a bin size of $\Delta_{\mmumu}=2\sqrt
2\sigma_r$ (which optimizes $S/\sqrt B$ for a flat background) for
comparing the $a$ signal to the above stipulated background.  As for
resolutions, it is stated in \cite{priceatlas} that the resolution at
the $J/\psi$ is around $54\mev$ while that at the $\upsi$ is close to
$170\mev$. We use a linear interpolation for other values of $\mmumu$.
Assuming $L=10\pbi$ of integrated luminosity, the background event
numbers $N_{\Delta {\mmumu}}$ in the intervals of size
$\Delta {\mmumu}=2\sqrt 2\sigma_r$ are $4055$ at $\ma=8\gev$, $50968$
at $\ma=\mupsi$ and $9620$ at $\ma=10.5\gev$.  We take the square root
to determine the $1\sigma$ fluctuation level.

We now consider the $a\to\mupmum$ signal rates. From
Fig.~\ref{totsigslhc}, we see that at $\tanb=10$ the total $a$ cross
section ranges from about $4.2\times 10^5\pb(\cta)^2\sim 4200\pb$ at
$\ma=8\gev$ to $\sim 8500\pb$ at $\ma\lsim 2m_B$ for $\rts=14\tev$.
The cross section for $a\to \mupmum$ assuming $\tanb=10$ and
$\cta=0.1$ will then range from $4200-8500\pb \times (\br(a\to
\mupmum)\sim 0.003)\sim 12-25\pb$. As discussed above, we will write
the total $a$ efficiency in the form $\effatlas=0.1 \times r$.
Multiplying the above cross section by $\effatlas$ and by the
$Erf(1)=0.8427$ acceptance factor for the ideal interval being
employed and using $L=10\pbi$ (as employed above in computing the
number of background events), we obtain $a$ event numbers of
$10\times r$, $18.5\times r$ and $21\times r$ at $\ma=8\gev$, $\mupsi$
and $10.5\gev$, respectively.  The statistical significances of the
$a$ peaks for $L=10\pbi$ are then $r\times$ the $r=1$ results of $0.16
\sigma$, $0.08 \sigma$ and $0.22 \sigma$, respectively.

Of course, we currently expect that substantial early running will
mostly take place at $\sqrt s=7\tev$ and $\sqrt s=10\tev$. As noted
earlier, lower $\rts$ implies a somewhat smaller $a$ cross section in
the $[8\gev,2m_B]$ mass interval on which we are focusing.  Roughly,
relative to $\rts=14\tev$, the $a$ cross section decreases by a factor
of $\sim 1.3$ at $\rts=10\tev$ and a factor of $\sim 1.7$ at
$\rts=7\tev$ in this mass interval. Since the backgrounds are also
basically $gg$ fusion induced, we presume that these same factors will
apply to them.  At $\rts=10\tev$ ($\rts=7\tev$) this then will reduce
the statistical significances given above by a factor of
$1/\sqrt{1.3}$ ($1/\sqrt{1.7}$). The statistical significances at
$\ma=8\gev$, $\mupsi$ and $10.5\gev$ are, respectively, then
$0.14\sigma$, $0.07\sigma$, $0.19\sigma$ at $10\tev$ and $0.12\sigma$,
$0.06\sigma$, $0.17\sigma$ at $7\tev$, all to be multiplied by $r$.  

\begin{table}
\caption{Luminosities ($\fbi$) needed for $5\sigma$ if $\tanb=10$ and $\cta=0.1$.}
\label{comparisontable2}
\begin{center}
\begin{tabular}{|c|c|c|c|}
\hline
Case  & $\ma=8\gev$ &$\ma=\mupsi$ & $\ma\lsim 2m_B$ \cr
\hline
ATLAS LHC7 & $17/r^2$ & $63/r^2$  &  $9/r^2$ \cr
ATLAS LHC10 & $13/r^2$ & $48/r^2$ & $7/r^2$ \cr
ATLAS LHC14 & $10/r^2$ & $37/r^2$ & $5.4/r^2$ \cr
\hline
\end{tabular}
\end{center}
\end{table}

Given the above results, we can tabulate the integrated luminosity $L$
needed to achieve a $5\sigma$ significance at each of the three
energies.  The results appear in Table~\ref{comparisontable2}.  The
required $L$'s away from the Upsilon resonance may be achieved after a
year or two of LHC operation.  The sensitivity of the required
luminosities to $r$ shows the importance of firmly establishing the
precise efficiencies for background and signal.  We look forward to
continued and detailed work by the ATLAS collaboration in this area.
Of course, we must not forget that the required $L$'s are very
sensitive to $\tanb$, $\cta$ and $\br(a\to \mupmum)$; very roughly for
$\tanb\neq 10$, $\cta\neq 0.1$ and/or $\br(a\to\mupmum)\neq 0.003$ the
tabulated luminosities need to be multiplied by
\beq 
\left({0.003\over \br(a\to\mupmum)}\right)^2\left({0.1\over
    \cta}\right)^4\left({10\over \tanb}\right)^{3.2-3.6}\,,  
\eeq 
where the $3.2$ applies for $\ma\sim 8\gev$ and the $3.6$ applies for
$\ma\lsim 2m_B$.  Depending upon the precise value of $\ma$ and
$\tanb$, in the $\ma$ mass range under discussion Fig.~\ref{bramumu}
shows that $\br(a\to \mupmum)$ can range from a low of $0.0023$ at
$\tanb=1.5$ and $\ma\lsim 2m_B$ to a high of $0.0033$ for $\tanb\geq
3$ and $\ma=8\gev$. The minimum values of $\cta$, with and without
placing a maximum on the light-$a$ finetuning measure $G$, were
detailed in Table~\ref{ctatable}.

Studies by CMS analogous to the ATLAS studies discussed above are
under way~\cite{bortprivatecom}.

\section{Conclusions}
\label{conclusions}

In this paper we have shown that a dedicated analysis of the dimuon
spectrum at the Tevatron and LHC at low masses, \ie\ $\mmumu\lsim
2m_B$, will provide very important constraints on models containing a
light CP-odd Higgs boson.  We employed the published $L=630\pbi$ CDF
analysis of the dimuon spectrum between $\sim 6.3\gev$ and $9\gev$ by
CDF and found that constraints on the $b\anti b a$ coupling $\cabb$
become competitive with those from $\Upsilon(nS)\to \gam a$ decays for
$8.5\gev\lsim \ma\lsim 9\gev$, and will be superior for larger data
sets.  In addition, only hadron colliders have the kinematic reach to
constrain $|\cabb|$ in the important region $\mupsiii\lsim\ma\lsim
2m_B$. In particular, for $L=10\fbi$, the Tevatron will provide
significant constraints on the $|\cabb|\gsim 1$ portion of the
$8\gev\lsim \ma\lsim 2m_B$ mass region that would allow an NMSSM ideal
Higgs scenario with an $\mh\sim 100-105\gev$ CP-even $h$ decaying
primarily via $h\to aa \to \tauptaum\tauptaum$ to be possible with
neither electroweak finetuning nor ``light-$a$'' finetuning.  It is
also very noteworthy that our rough estimates of the limits that CDF
could place on $|\cabb|$ using the $L=630\fbi$ event rates in the
$9\leq\ma\leq 12\gev$ region are such that the observed $a_\mu$
discrepancy could not be explained by a light $a$.

For the LHC, we have obtained rough estimates of what will be possible
using information available from the ATLAS collaboration, in
particular regarding the efficiency (for triggering, tracking, $p_T$
cuts, etc.) for retaining $a\to \mupmum$ events.  We find that it will
be possible to obtain a $5\sigma$ signal for a light $a$ with
$\tanb\sim 10$ and $\cta\sim 0.1$ throughout the entire range
$8\gev\lsim \ma\lsim 2m_B$ away from the Upsilon peaks for $L\sim
13\fbi$ at $\rts=10\tev$ or $L\sim 17\fbi$ at $\rts=7\tev$.  For
example, at $\ma=10.5\gev$, only $L\sim 7\fbi$ at $\rts=10\tev$ or
$L\sim 9\fbi$ at $\rts=7\tev$ is required to achieve a $5\sigma$
signal for such an $a$.

Of course, not all acceptable NMSSM models have $|\cabb|$ as large as
$\sim 1$. As an extreme example, at $\tanb=1.7$, $\cta\sim 0.1$ is
possible for small light-$a$ finetuning (corresponding to $\cabb\sim
0.17$).  In this case, the $a$ cross section at $\ma\lsim 2m_B$ is
about a factor of 18 smaller than at $\tanb=10$ and $\cta\sim 0.1$.
Using statistical extrapolation this suggests that as much as $324$
times more luminosity would be needed to achieve the same statistical
significances as above.  However, one should keep in mind that it may
in the end be possible to obtain net efficiencies for the $a$ at ATLAS
and CMS in excess of the current ATLAS estimate of $10\%$.  Indeed,
early CMS studies suggest that net efficiencies might be as high as
$30\%$~\cite{bortprivatecom}.  Since the needed $L$ scales inversely
with the square of the efficiency, assuming $\rts=14\tev$ and $r=3$
one finds that a $5\sigma$ signal could be achieved for $\tanb=1.7$
and $\cta\sim 0.1$ with $L\sim 195\fbi$, an integrated luminosity that
should be achieved in the not too distant future, although background levels might be
larger at the higher instantaneous luminosities needed to achieve such
large total $L$.

Overall, this kind of search is quite important given that there are
many models in which light $a$'s are present that have significant,
even if not enhanced, couplings to gluons via quark loops and that
would have reasonable $a\to \mupmum$ branching ratio. Searching for
such $a$'s and constraining their possible masses and couplings is an
important general goal and both Tevatron and LHC data will be of great
value.

\acknowledgments 

We thank D. Price, D.  Bortoletto, H. Evans, Z. Gecse, C. Mariotti and
M. Pellicioni for helpful conversations regarding the ATLAS and CMS
analyses, respectively.  We particularly want to thank Y. Yang for
performing the Monte Carlo study needed to estimate the net ATLAS
efficiency.  JFG was supported by U.S. DOE grant No.
DE-FG03-91ER40674, the National Science Foundation under Grant No.
PHY05-51164 while at KITP, the Aspen Center for Physics and, during
the completion of the work, as a Scientific Associate at CERN.

\end{document}